\documentclass[
  aps, prb,
  reprint,
  amsmath,
  floatfix,
  ]{revtex4-2}

\usepackage[T1]{fontenc}
\usepackage{libertinus}
\usepackage{libertinust1math}
\usepackage[cal=stixtwoplain]{mathalpha}
\usepackage{booktabs}
\usepackage{csquotes}
\usepackage{enumitem}

\usepackage{microtype}
\usepackage{bm}

\usepackage{xcolor}
\usepackage{graphicx}

\usepackage{siunitx}
\usepackage[version=4]{mhchem}
\usepackage{mleftright}
\mleftright

\usepackage{hyperref}
\colorlet{hypercolor}{blue!50!black}
\hypersetup{
  colorlinks,
  linkcolor=hypercolor,
  filecolor=hypercolor,
  citecolor=hypercolor,
  urlcolor=hypercolor,
}

\makeatletter
  \patchcmd\frontmatter@setup{\normalfont}{\normalfont\sffamily}{}{}
  \patchcmd\frontmatter@title@format{\bfseries}{\sffamily\bfseries}{}{}
  \patchcmd\frontmatter@affiliationfont{\it}{\slshape}
  \patchcmd\frontmatter@title@format{\large}{\Large}{}{}
  \patchcmd\section{\bfseries}{\sffamily\bfseries}{}{}
  \patchcmd\subsection{\bfseries}{\sffamily\bfseries}{}{}
  \patchcmd\subsubsection{\itshape}{\sffamily}{}{}
  \patchcmd\@makecaption{\rmfamily}{\sffamily}{}{}
  \patchcmd\@makecaption{\rmfamily}{\sffamily\bfseries}{}{}
\makeatother

\makeatletter
  \def\bibsection{%
    \expandafter\section\expandafter*\expandafter{\refname}%
    \@nobreaktrue
  }%
\makeatother

\def\e{\mathrm{e}}
\def\i{\mathrm{i}}

\let\vec\bm

\makeatletter
  \pretocmd\@make@capt@title{\begingroup\bfseries}{}{}
  \patchcmd\@make@capt@title{\@caption@fignum@sep}{\@caption@fignum@sep\endgroup}{}{}
  \def\@caption@fignum@sep{: }

\makeatother

\NewDocumentCommand\titlecaption{ m m }{%
  \caption[#1]{\textbf{#1}\hskip 1.5\fontdimen2\font plus 1em minus 1.5\fontdimen4\font\relax#2}%
}
\NewDocumentCommand\sublabel{ m }{(#1)}

\NewDocumentCommand\lisaplus{}{LISA\!\textsuperscript{\bfseries +}}

\DeclareSIUnit\ions{ions}
\begin{document}

\title{Dually tunable YBCO coplanar waveguide resonators \\ based on helium-ion-generated Josephson inductances}

\def\pitaffil{%
	\affiliation{Physikalisches Institut, Center for Quantum Science~(CQ) and \lisaplus, Universität Tübingen, 72076~Tübingen, Germany}
}
\author{Kenny~Fohmann}
\email{kenny.fohmann@uni-tuebingen.de}
\pitaffil
\author{Timo~J.~Glebe-Märklin}
\pitaffil
\author{Benedikt~Wilde}
\pitaffil
\author{Mohamad~Kazouini}
\pitaffil
\author{Christoph~Schmid}
\pitaffil
\author{Dieter~Koelle}
\pitaffil
\author{Reinhold~Kleiner}
\pitaffil
\author{Daniel~Bothner}
\email{daniel.bothner@uni-tuebingen.de}
\pitaffil

\begin{abstract}
Superconducting microwave circuits with and without Josephson inductances are the Swiss Army knife for many experiments and technologies from quantum information science to astrophysical particle detectors.
Despite a large variety of existing circuit types, thin film materials and Josephson junction technologies, a flexible and reliable platform for high-magnetic-field and high-temperature applications is yet to be found.
In this manuscript, we investigate coplanar waveguide cavities made of the high-temperature cuprate superconductor YBa$_2$Cu$_3$O$_7$~(YBCO), integrated with Josephson inductances and quantum interferometers that are generated by the controlled local irradiation of the YBCO with a focused helium ion beam.
We obtain strongly flux-tunable microwave resonators, which not only display periodic interferometer oscillations of resonance frequency and decay rate, but also a superimposed Fraunhofer-like modulation pattern.
The latter originates from the magnetic field tuning of the individual Josephson junction critical currents due to the out-of-plane junction barriers.
It allows adjusting resonance frequency and flux responsivity independently of each other, potentially enabling tunable microwave circuits with low sensitivity to external magnetic-field noise over a broad range of frequencies.
Finally, we investigate the temperature dependence of the cavities, show that they have promising characteristics up to $\qty{14}{\kelvin}$, and present a model for the junction-induced cavity losses.
\end{abstract}
\maketitle

\section*{Introduction}
\vspace{-1mm}
Superconducting microwave circuits belong to the most important and versatile components in numerous modern technologies.
They are key in quantum information science~\cite{Krantz2019, Blais2021, Arute2019, Kim2023}, quantum-limited signal processing~\cite{CastellanosBeltran2008, Bergeal2010, Macklin2015}, and quantum simulation~\cite{Houck2012, Carusotto2020}; they provide sensing and control tools for micromechanical oscillators~\cite{Teufel2011, Youssefi2022, Bild2023}, low-frequency photons~\cite{Eichler2018, Bothner2021, NajeraSantos2024}, spins~\cite{Bienfait2016, Wang2023} or magnons~\cite{LachanceQuirion2020, Golovchanskiy2021, Kounalakis2022}; they constitute highly sensitive particle detectors~\cite{Day2003, Endo2019, Ulbricht2021}, and they are essential for different types of hybrid quantum systems with quantum dots~\cite{Samkharadze2018, Mi2018}, spin ensembles~\cite{Ghirri2015, Albanese2020, VelluirePellat2025}, or ultracold atoms~\cite{Kaiser2022, Kumar2023}.
Lately, there is considerable interest in both linear and nonlinear microwave circuits that are based on materials other than the standard choice, aluminum, in particular materials that promise to offer a much larger operating regime, such as access to higher temperatures or compatibility with large external magnetic fields.
Higher temperatures than \unit{\milli\kelvin} are interesting for low-cost setups and higher cooling power~\cite{Vaartjes2024, Anferov2025}, for space missions~\cite{Griffin2016, Baselmans2017, Chakrabarty2019}, multifrequency sensing of phase transitions~\cite{Miksch2021}, and for hybrid systems in which it is hard to efficiently cool down the superconducting component to~\unit{\milli\kelvin}~\cite{Kumar2023}.
Large magnetic fields on the other hand are important for hybrid systems with phonons, magnons, topological qubits or spins~\cite{Shevchuk2017, Rodrigues2019, Zoepfl2020, Schmidt2020, Schmidt2024, LachanceQuirion2020, Golovchanskiy2021, Kounalakis2022, Samkharadze2018, Mi2018, Ghirri2015, Albanese2020, Hyart2013, Plugge2017}, and potentially even towards the search for dark matter axions~\cite{Bartram2023, DiVora2023, Chaudhuri2019, Kuenstner2025}.
In the linear domain, niobium, niobium-alloys or the high-temperature superconductor YBa$_2$Cu$_3$O$_7$~(YBCO) can enable low-loss circuits in the several kelvin and several tesla regime~\cite{Ghirri2015, Samkharadze2016, Kwon2018, Roitman2023}.
The inclusion of Josephson inductances for controllable nonlinearities, \textit{in-situ} tunability, or flux sensitivity has proven much more challenging with these materials, though.
Possible solutions that have been reported include the use of high-kinetic-inductance nonlinearities instead of Josephson junctions (JJs)~\cite{Eom2012, Parker2022, Frasca2024}, nano-constrictions and nanowires as nonlinear elements~\cite{Tholen2009, Kennedy2019, Xu2023, Uhl2024}, or more unconventional technologies with e.g.~semiconducting or graphene-based junctions~\cite{Luthi2018, PitaVidal2020, Kringhoj2021, Butseraen2022}.
Each of these technologies has its own limitations though, such as a restricted range of critical currents in constrictions or low Kerr nonlinearities and low tunability when using kinetic inductance.
For YBCO, another alternative is available: Barrier Josephson junctions (bJJs) based on nanometric irradiation with a focused helium ion beam (He-FIB) \cite{Cybart2015, Mueller2019}, which locally changes the crystal structure by displacing oxygen atoms and induces narrow normal conducting or insulating barriers~\cite{Zaluzhnyy2024}.
Compared to other JJ technologies in YBCO such as step-junctions or grain boundary junctions, the He-FIB~bJJs (or short He-bJJs) do not require specific substrates and are extremely flexible with respect to their critical current density~\cite{Mueller2019}, which can be easily varied over orders of magnitude on a single chip by locally choosing the ion dose.
The only implementation of YBCO microwave circuits with He-bJJs to date is based on lumped element LC circuits~\cite{Uhl2023} and has revealed a high tunability of the resonance frequency up to GHz.
While the reported devices are promising for experiments like dispersive magnetometry~\cite{Hatridge2011, LevensonFalk2016} or first hybrid devices with phonons and magnons, they also leave considerable room for improvements and call for further experiments.
For instance, the impact of the resonator layout on decay rates and tuning range, the circuit temperature dependence, or a possible Fraunhofer-modulation due to the individual He-bJJs, which in contrast to most other microwave-integrated JJs have their barrier oriented perpendicular to the substrate, have not been investigated yet.
\begin{figure*}
	\includegraphics{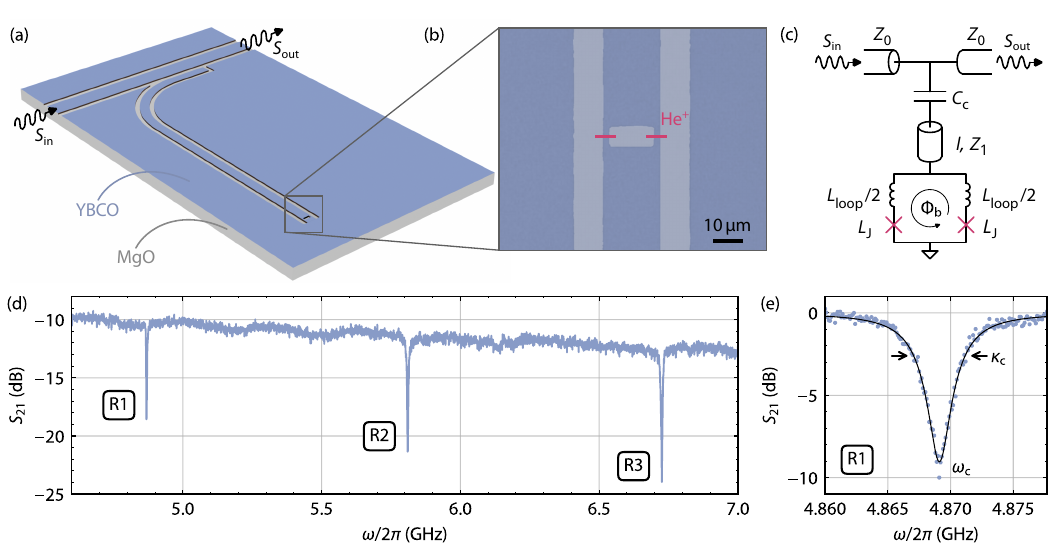}
	\vspace{-1.4mm}
	\titlecaption{YBCO coplanar waveguide cavities for integration with helium-ion-generated Josephson inductances.}{%
		\sublabel{a}~Schematic of a quarter-wave coplanar waveguide (CPW) cavity as used in this work (not to scale), capacitively side-coupled to a CPW transmission line. The substrate material is MgO, the superconducting film is $\qty{50}{\nano\meter}$ thick M-type YBCO. For capacitive coupling to the feedline, the cavity has a coupling elbow near its open end. Close to its shorted end far from the feedline, the cavity contains a loop in the CPW center conductor for the SQUID. Characterization is performed by measurements of the scattering matrix element $S_{21} = S_\mathrm{out}/S_\mathrm{in}$.
		\sublabel{b}~False-color optical microscopy image around the SQUID loop of a typical sample. YBCO is blue, MgO substrate is gray, the two pink lines indicate the position of the irradiation with helium ions to generate two Josephson junctions~(JJs).
		\sublabel{c}~Circuit equivalent of a capacitively side-coupled transmission line cavity with an integrated SQUID at its shorted end. The feedline has a characteristic impedance $Z_0$, the cavity itself has a length $l$ and a characteristic impedance $Z_1$. The coupling capacitor is $C_\mathrm{c}$. The SQUID is characterized by a loop self-inductance $L_\mathrm{loop}$ and two Josephson junctions, each with an inductance $L_\mathrm{J}$ that can be tuned by threading an external magnetic flux $\Phi_\mathrm{b}$ into the loop.
		\sublabel{d}~Transmission~$S_{21}$ along the feedline with three side-coupled YBCO CPW cavities of different length (labeled R1 to R3) before bJJ writing. The resonance frequencies are nearly equidistantly distributed between $\omega_\mathrm{c1} = 2\pi\times \qty{4.869}{\giga\hertz}$ and $\omega_\mathrm{c3} = 2\pi\times \qty{6.727}{\giga\hertz}$, the linewidths are on the order of $\kappa_\mathrm{c} \sim 2\pi\times \qty{5}{\mega\hertz}$. All values are detailed in Appendix\ref{app:dev_parameters}.
		\sublabel{e}~Zoom into the resonance of R1 after background correction. Symbols are data, line is a fit. R1 has internal and external quality factors $Q_\mathrm{int, c} \approx 3350$ and $Q_\mathrm{ext, c} \approx 1800$, respectively, which correspond to internal and external decay rates $\kappa_\mathrm{int, c} \approx 2\pi\times \qty{1.454}{\mega\hertz}$ and $\kappa_\mathrm{ext, c} \approx 2\pi\times \qty{2.731}{\mega\hertz}$.
	}
	\vspace{-0.85mm}
	\label{fig:figure1}
\end{figure*}
Here, we present the realization and experimental characterization of YBCO coplanar waveguide resonators with integrated He-bJJs.
As base material we use M-type YBCO from the company \textit{ceraco ceramic coating GmbH}~(Ceraco).
As a first experiment we confirm that, despite its different stoichiometry and structure from the previously used film types, high-quality He-bJJs can be achieved in M-type YBCO.
The resulting junction critical current densities have a similar ion-dose dependence as reported for other YBCO films.
In a next step, we introduce low critical-current He-bJJs (${\sim}\qty{2}{\micro\ampere}$) into pre-patterend quarter-wave coplanar waveguide cavities.
The bJJs are integrated in pairs of two to form a superconducting quantum interference device~(SQUID), i.e.,~a magnetic-flux-tunable Josephson inductance, and we find a resulting flux-tunability of the cavity resonance frequencies up to the GHz range.
For SQUID flux values beyond a few flux quanta, a Fraunhofer-like envelope of the SQUID modulations is revealed, an intriguing observation, which enables choosing flux-responsivity and resonance frequency independently of each other, and allows developing magnetic-field-tunable resonators based on single JJs.
Furthermore, we present a simple model to describe the bJJ-induced energy losses in the cavities and discuss strategies for their minimization.
Finally, we characterize the temperature-dependence of the SQUID cavities in a temperature range between $\qty{2.1}{\kelvin}$ and $\qty{14}{\kelvin}$ and observe large tunabilities of the cavity resonance frequencies up to the highest investigated temperature.
Our results pave the way for microwave SQUID-circuit experiments at elevated temperatures, for YBCO-based Josephson parametric amplifiers, for hybrid systems with micromechanical oscillators and magnons in large magnetic fields, and for flux-noise-insensitive tunable microwave cavities.
\vspace{-3mm}
\section*{Results}
\vspace{-2mm}
\subsection*{Devices and setup}
\vspace{-0.5mm}
The base layout for the superconducting microwave resonators are coplanar waveguide (CPW) transmission line cavities, side-coupled to a CPW feedline by means of a coupling capacitance $C_\mathrm{c}\approx \qty{18}{\femto\farad}$, cf.~Fig.~\ref{fig:figure1}.
We investigate three of those cavities here, all coupled to a single common feedline with characteristic impedance $Z_0 \approx \qty{50}{\ohm}$.
The cavities have a center conductor width $S = \qty{19.5}{\micro\meter}$ and two gaps to ground with width $W = \qty{9.5}{\micro\meter}$, leading to a characteristic impedance $Z_1 \approx \qty{60}{\ohm}$.
At their end close to the feedline the cavities are terminated by an open end, and at their end far from the feedline by a short to ground, i.e., they are of the quarter-wave ($\lambda/4$) hanger-type.
Close to the shorted end, the CPW center conductor contains a $\qty{14.5}{\micro\meter}\times \qty{6.5}{\micro\meter}$ large hole, cf.~Fig.~\ref{fig:figure1}, that will form the SQUID after helium-ion irradiation in a later step.
The YBCO microbridges on both sides of the SQUID loop, where the Josephson junctions will be placed, have a width $a = \qty{2.5}{\micro\meter}$.
The total lengths of the three cavities, labeled R1, R2, and R3 throughout this manuscript, are roughly equidistantly distributed between $l_\mathrm{R1} = \qty{5.772}{\milli\meter}$ and $l_\mathrm{R3} = \qty{4.141}{\milli\meter}$, leading to fundamental-mode resonance frequencies between $\omega_\mathrm{c1} = 2\pi\times \qty{4.869}{\giga\hertz}$ and $\omega_\mathrm{c3} = 2\pi\times \qty{6.727}{\giga\hertz}$ before writing of the bJJs.
The fabrication of the chip is described in Appendix\ref{app:devicefab}, and the values for all cavity parameters can be found in Appendix\ref{app:dev_parameters}.
After base circuit fabrication, but before helium-ion irradiation, the cavities are spectroscopically pre-characterized.
To this end, the chip is connected to two coaxial cables via a printed circuit board (PCB), which route microwave signals from a vector network analyzer (VNA) to the on-chip feedline and back, i.e., we perform a measurement of the complex scattering matrix element $S_{21} = S_\mathrm{out}/S_\mathrm{in}$.
All experiments, except for the temperature-dependence discussed in the last part of the manuscript, have been conducted with the sample immersed in liquid helium, i.e., at a measurement temperature $T_\mathrm{s} = \qty{4.2}{\kelvin}$.
Details on the measurement setup can be found in Appendix\ref{app:setups}.
The result of the pre-characterization is displayed in Fig.~\ref{fig:figure1}(d,~e).
Three clear and sharp resonances can be identified in the broadband transmission, and from fitting the data around each of the resonances we obtain the resonance frequencies $\omega_\mathrm{c}$ as well as the total and external linewidths $\kappa_\mathrm{c}$ and $\kappa_\mathrm{ext, c}$, respectively.
From a combination of the length $l$ of each cavity, its resonance frequency $\omega_\mathrm{c}$, its external linewidth $\kappa_\mathrm{ext, c}$ and the calculated capacitance per unit length $C' = \qty{147}{\pico\farad\per\meter}$, we can finally determine the inductance per unit length $L' \approx \qty{518}{\nano\henry\per\meter}$ of the cavity CPW and subsequently the loop self-inductance $L_\mathrm{loop}\approx \qty{36}{\pico\henry}$, which are important parameters to quantify the properties of the He-FIB-induced bJJs in a later experimental stage.
The theoretical model of the cavities, a description of the data processing and fitting routines, and the values for all frequencies, linewidths, capacitances and inductances can be found in Appendices\ref{app:fitting}~to\ref{app:dev_parameters}.
Notably, the internal quality factors $Q_\mathrm{int, c} = \omega_\mathrm{c}/\kappa_\mathrm{int, c}\sim3000$ of the CPW cavities here ($\kappa_\mathrm{int, c} = \kappa_\mathrm{c} - \kappa_\mathrm{ext, c}$) are significantly larger than the ones of the lumped element circuits in Ref.~\cite{Uhl2023}, almost by one order of magnitude.
We attribute this to an improved fabrication method in combination with lower resonance frequencies and the larger width of the current-carrying resonator parts; in the current work the main current-path is along the $\qty{19.5}{\micro\meter}$ wide CPW center conductor (and the ground planes), while in Ref.~\cite{Uhl2023} the width of the superconducting lines was only ${\sim}\qty{1.5}{\micro\meter}$.
In earlier works with even wider center conductors \cite{VelluirePellat2023, Ghirri2015} internal quality factors $10^4 - 5\cdot 10^4$ could be achieved, albeit in those studies also thicker YBCO films were used, which increases the quality factors further, but which is currently not compatible with He-bJJs.
Before we integrate suitable Josephson junctions into the three cavities, we investigate the properties of individual YBCO bJJs by means of electric transport measurements and determine the helium-ion-dose dependence of their characteristics, which has not been studied for M-type films yet.
\vspace{-3mm}
\subsection*{DC characteristics and ion-dose dependence of He-FIB bJJs}
%
\begin{figure}
	\includegraphics{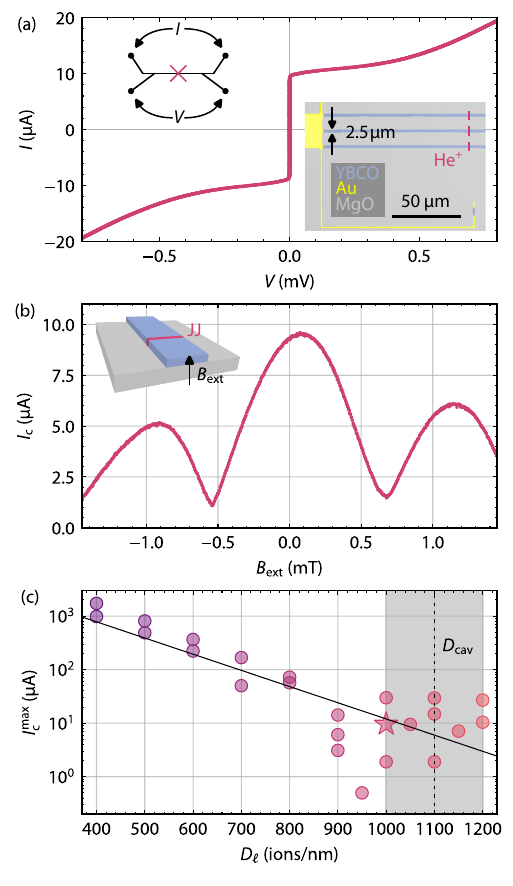}
	\vspace{-8mm}
	\titlecaption{Transport characteristics and critical-current ion-dose dependence of M-type YBCO He-bJJs.}{%
		\sublabel{a}~IVC of a He-FIB bJJ with an ion line dose of $D_\ell = \qty[per-mode=symbol]{1000}{\ions\per\nano\meter}$. Critical current is $I_\mathrm{c} \approx \qty{9.5}{\micro\ampere}$. Top-left inset shows a schematic of the four-terminal IV measurement. Bottom-right inset shows a false-color optical image of the $a_\mathrm{dc} = \qty{2.5}{\micro\meter}$ wide bridge structures used for these experiments, depicting one set of three parallel bridges that are connected to a common ground but biased and measured individually.
		\sublabel{b}~Critical junction current $I_\mathrm{c}$ vs applied magnetic field $B_\mathrm{ext}$ for the junction of panel~\sublabel{a}. Inset schematically shows the application of $B_\mathrm{ext}$ perpendicular to the chip surface, i.e., parallel to the JJ barrier-plane. The Fraunhofer-like modulation of $I_\mathrm{c}(B_\mathrm{ext})$ down to $I_\mathrm{c}^\mathrm{min} < \qty{2}{\micro\ampere}$ indicates a junction barrier with decent uniformity; small imperfections (cf.~main text for more details) lead to the remaining asymmetry.
		\sublabel{c}~Maximum critical currents $I_\mathrm{c}^\mathrm{max}$ vs ion dose of $24$ helium-ion bJJs. The critical current decreases exponentially with the line dose $D_\ell$; symbols are data extracted from $I_\mathrm{c}(B_\mathrm{ext})$ datasets, solid line is an exponential fit with the characteristic dose $D_0 = \qty[per-mode=symbol]{143}{\ions\per\nano\meter}$ as a fit parameter. The shaded area indicates the desired region for highly-tunable but weakly-nonlinear microwave circuits, the dashed line indicates the dose $D_\mathrm{cav}$ chosen for the microwave devices in this work, and the star-shaped symbol corresponds to the bJJ discussed in panels \sublabel{a} and \sublabel{b}.
	}
	\label{fig:figure2}
\end{figure}
For the determination of the critical junction current as a function of the ion line dose $D_\ell$ using transport experiments, we prepared a $\qty{10}{\milli\meter}\times\qty{10}{\milli\meter}$ large chip from the same wafer as the microwave chip discussed above, and patterned it with $96$~microbridges, each of which can be individually characterized by means of a four-terminal current-voltage-characteristic~(IVC).
The width of the microbridges $a_\mathrm{dc}$ is approximately identical to the SQUID arm width in the microwave devices, i.e., $a_\mathrm{dc} \approx a = \qty{2.5}{\micro\meter}$.
The fabrication of the dc chip is detailed in Appendix\ref{app:devicefab}, and is essentially identical to that of the microwave chip.
For the dose-series, we direct-wrote He-bJJs into $24$ of the $96$ available microbridges with nominal line doses between $D_\ell^\mathrm{min} = \qty[per-mode=symbol]{400}{\ions\per\nano\meter}$ and $D_\ell^\mathrm{max} = \qty[per-mode=symbol]{1200}{\ions\per\nano\meter}$.
In a first irradiation session with \num{18}~bJJs we chose a dose step size of $\Delta D_\ell = \qty[per-mode=symbol]{100}{\ions\per\nano\meter}$ and generated two junctions per dose value.
In an additional later irradiation session, we wrote six more junctions in the most interesting regime $D_\ell\ge \qty[per-mode=symbol]{900}{\ions\per\nano\meter}$ with a dose step size of $\qty[per-mode=symbol]{50}{\ions\per\nano\meter}$.
Using a low-noise experimental setup for transport measurements in liquid helium, cf.~Appendix\ref{app:setups}, we then characterized the junction IVCs as well as the critical junction current $I_\mathrm{c}$ as a function of an externally applied magnetic field perpendicular to the chip surface $B_\mathrm{ext}$.
The critical current $I_\mathrm{c}$ here is defined as that dc current at which we detect a dc voltage drop across a JJ of $V_\mathrm{dc} \approx \qty{3}{\micro\volt}$~(for six JJs $\qty{4}{\micro\volt}$ and for one JJ $\qty{5}{\micro\volt}$ due to slightly increased noise), which is a few times the root-mean-square noise-voltage of the used electronics.
Hence, for small $I_\mathrm{c} \lesssim \qty{5}{\micro\ampere}$, the detected critical current $I_\mathrm{c}$ might be reduced by thermal effects and a corresponding rounding of the IVCs as compared to the ideal, noiseless critical current $I_0$ , i.e.,~$I_\mathrm{c} \lesssim I_0$ (thermal noise current is $I_\mathrm{th} = 2\pi k_\mathrm{B} T_\mathrm{s}/\Phi_0 \approx \qty{176}{\nano\ampere}$ at $\qty{4.2}{\kelvin}$).
Our main results are summarized in Fig.~\ref{fig:figure2}.
Similar to the results of earlier experiments using S-type YBCO films from Ceraco~\cite{Cybart2015} or YBCO films grown by pulsed laser deposition \cite{Mueller2019}, we observe for most junctions non-hysteretic Josephson-like IVCs, that seem to be well-described by a resistively-shunted junction (RSJ) model with a clear transition from the superconducting branch to the voltage state.
In Fig.~\ref{fig:figure2}\sublabel{a} the IVC of one of the higher-dose bJJs is shown ($D_\ell = \qty[per-mode=symbol]{1000}{\ions\per\nano\meter}$) with a critical current $I_\mathrm{c} \approx \qty{9.5}{\micro\ampere}$ (and a slight thermal smearing around $I_\mathrm{c}$), a regime that is very favorable for highly tunable, low-nonlinearity microwave devices.
In particular the higher-dose bJJs also show a pronounced modulation of the critical current with the external magnetic field, which resembles a Fraunhofer-like interference pattern and indicates a reasonably good uniformity of the helium-ion-induced barrier, cf.~Fig.~\ref{fig:figure2}\sublabel{b}, once again similar to earlier results in other YBCO film types~\cite{Mueller2019}.
Deviations from an ideal Fraunhofer interference pattern such as higher-than-expected side domes, asymmetries or minima at non-integer multiples of the first minimum can be attributed to inhomogeneous critical current densities~\cite{Dynes1971, LeFebvre2022}, inhomogeneous magnetic fields~\cite{Miller1985, Rosenthal1991}, and the junction being close to the nonlocal Josephson regime~\cite{Moshe2008, Clem2010, Boris2013}.
For the junction geometry realized here, the theoretical prediction for the magnetic field at the first minimum is $B_\mathrm{m} \approx 1.5\Phi_0/a_\mathrm{dc}^2 \approx \qty{0.5}{\milli\tesla}$~\cite{Clem2010}, which agrees reasonably well with the experimental findings, cf.~Fig.~\ref{fig:figure2}\sublabel{b}.
When comparing the maximum critical currents obtained from the $I_\mathrm{c}(B_\mathrm{ext})$ data for the different ion doses $D_\ell$,  cf.~Fig.~\ref{fig:figure2}\sublabel{c}, we find that $I_\mathrm{c}(D_\ell)$ can be varied by roughly three orders of magnitude in our moderate dose range.
The critical currents decrease exponentially with $D_\ell$ and from a fit to the expression $I_\mathrm{c}(D_\ell) = I_{\mathrm{c}0}e^{-D_\ell/D_0}$, we obtain the characteristic dose $D_0 = \qty[per-mode=symbol]{143}{\ions\per\nano\meter}$ in good agreement with earlier results on MgO substrate~\cite{Mueller2019}.
The additional fit parameter, the critical current without He-FIB irradiation $I_{\mathrm{c}0} \sim \qty{12.6}{\milli\ampere}$, describes the extrapolated critical current of the junctionless microbridge, but has no direct relevance for the experiments here.
A second important observation is that the critical currents scatter considerably around the fit line, especially for the higher doses.
This is an intrinsic challenge of He-FIB bJJs and has also been observed using other film types and substrates, where the critical current densities varied by more than an order of magnitude for a single dose~\cite{Mueller2019, Proepper2025}.
The exact origin of these variations is not completely understood yet, but likely contributions are from inhomogeneous film characteristics, variations in ion-beam focus size from one JJ to the next, vibrations during irradiation or inaccuracies in dose calibration~(undetected beam current fluctuations).
A power-law scaling of all characteristic junction voltages $V_\mathrm{c} = I_\mathrm{c}R_\mathrm{n} \propto j_\mathrm{c}^\alpha$ with the critical current densities $j_\mathrm{c}$ and the voltage-state resistances $R_\mathrm{n}$, however, indicates that all devices are still described by Josephson physics, cf.~Appendix\ref{app:add_data} and Fig.~\ref{fig:figure7}.
Wrapping up this dc excursion, we conclude that helium-ion-induced Josephson junctions can be realized in M-type YBCO with an ion-dose dependence very similar to the one found in previous studies using other YBCO film types and substrates.
From the dose series, we furthermore obtained a guideline for the doses required for specific Josephson microwave circuits, the exact value depending on the target device in mind.
Some of the most exciting applications of Josephson circuits like parametric amplifiers, optomechanical systems or dispersive magnetometers typically require $I_0$ in the few \unit{\micro\ampere} regime to guarantee large frequency-tunability, large flux responsivity, and a high dynamic range, and hence we chose to pattern our three microwave cavities with $D_\mathrm{cav} = \qty[per-mode=symbol]{1100}{\ions\per\nano\meter}$.
In future experiments it will also be interesting to test whether we can reliably achieve even lower critical currents as needed for instance for superconducting qubits (${\sim} 10$\textendash$\qty{100}{\nano\ampere}$) and to investigate the junction properties in the low-$I_0$ regime at \unit{\milli\kelvin} temperatures, where thermal currents should be negligible.
\clearpage
%
\subsection*{Flux-tuning of CPW cavities with He-FIB bJJ SQUIDs}
%
\begin{figure*}
	\includegraphics{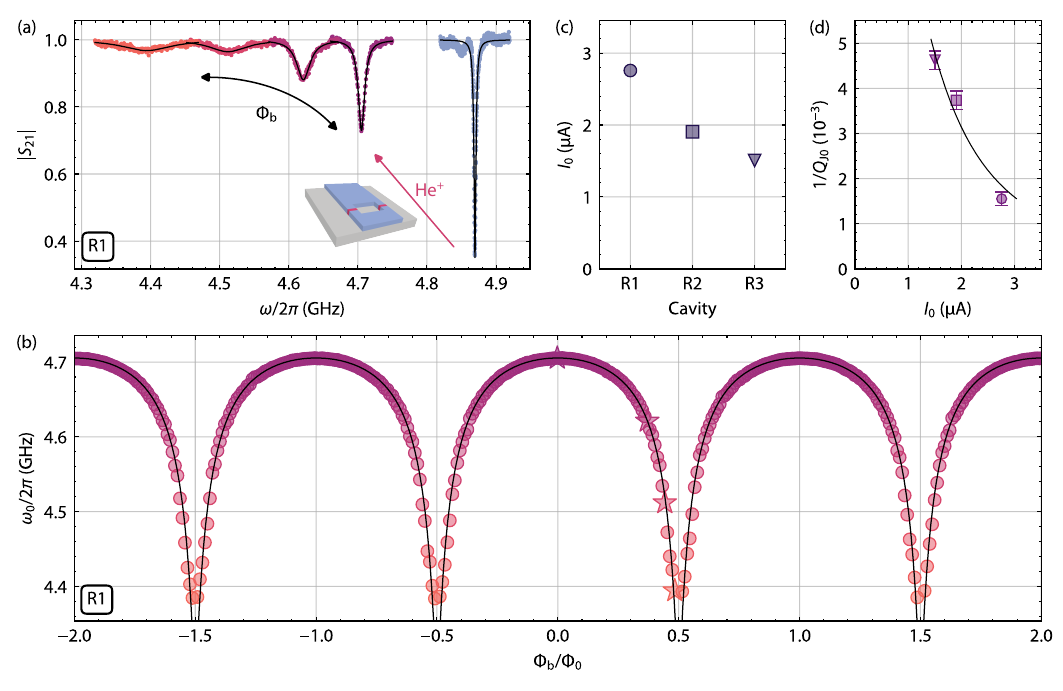}
	\vspace{-7mm}
	\titlecaption{Flux-tuning of a helium-ion-generated YBCO SQUID-cavity and parameters of the integrated Josephson junctions.}{%
		\sublabel{a}~Transmission $|S_{21}|$ around the resonance frequency of R1 without (light blue) and with (purple, pink, red, orange) He-bJJs for different values of bias flux $\Phi_\mathrm{b}$ through the SQUID. Introducing the junctions shifts the resonance frequency from $\omega_\mathrm{c} = 2\pi \times\qty{4.869}{\giga\hertz}$ to $\omega_0 = 2\pi\times \qty{4.705}{\giga\hertz}$, mainly due to the additional inductance of the JJs. When a magnetic field perpendicular to the chip surface is applied, equivalent to a bias magnetic flux $\Phi_\mathrm{b}$ through the SQUID, the resonance dip shifts further to lower values, here down to $\omega_0 = 2\pi\times\qty{4.39}{\giga\hertz}$ at $\Phi_\mathrm{b}/\Phi_0 \sim 0.48$. Symbols are data, black lines are fits, from which we extract for each $\Phi_\mathrm{b}$ the resonance frequency $\omega_0$ and the total and external cavity linewidth. For larger ranges of magnetic flux, we find a periodic modulation of the cavity parameters with period $\Phi_0$; the modulation of $\omega_0$ is shown in \sublabel{b} for R1. Symbols are extracted from fits, the four star symbols correspond to the four datasets in \sublabel{a}. From a fit to the periodic flux arcs, overlaid to the R1 data as black lines (analogous data/fits for R2 and R3 are shown in Appendix\ref{app:add_data}), we extract the (mean) critical current of a single junction $I_0 = \qty{2.76}{\micro\ampere}$.
		\sublabel{c}~Single-junction critical currents $I_0$ for all three cavities R1, R2, and R3 as obtained from the flux arc fits. All three are in the single-digit \unit{\micro\ampere} regime, which also correspond to small SQUID screening parameters $\beta_L = 2L_\mathrm{loop}I_0/\Phi_0 < 0.1$ with $L_\mathrm{loop} = \qty{36}{\pico\henry}$.
		\sublabel{d}~Sweetspot JJ loss factor $1/Q_{\mathrm{J}0} = 1/Q_{\mathrm{int}0} - 1/Q_\mathrm{int, c}$ vs critical junction current $I_0$, where $Q_{\mathrm{int}0}$ is the internal quality factor after junction writing at the flux sweetspot, and $Q_\mathrm{int, c}$ is the internal quality factor before JJ writing. The losses induced by the presence of the junctions decrease with increasing critical current; line is a fit ${\propto} I_0^{-2}$, and error bars consider the uncertainty in~$Q_\mathrm{int, c}$, cf.~Appendix\ref{app:errors}.
	}
	\vspace{-1.75mm}
	\label{fig:figure3}
\end{figure*}
After completing the pre-characterization of the microwave cavities and determining the desired ion dose, the writing of the Josephson junctions into the microwave SQUIDs is carried out.
As mentioned above, we chose an identical dose of $D_\mathrm{cav} = \qty[per-mode=symbol]{1100}{\ions\per\nano\meter}$ for all three cavities, and we expect a critical-current variation on the order of at least a factor $2$\textendash$3$ for nominally identical junctions at different locations on the chip.
After the He-FIB process, the chip is mounted into a low-noise microwave dipstick, and we add a small electromagnetic coil for the application of a magnetic field perpendicular to the chip surface.
Compared to the pre-characterization experiments, the setup is slightly more sophisticated now, cf.~Appendix\ref{app:setups}.
The coaxial input line is attenuated with $\qty{30}{\decibel}$, and the coaxial output line is equipped with a cryogenic high-electron-mobility transistor (HEMT) amplifier with ${\sim}\qty{38}{\decibel}$ of gain.
The HEMT amplifier is required for improved signal-to-noise ratios, since cavities with Josephson junctions are highly nonlinear and need to be probed with very small VNA powers.
To guarantee that the cavities are always probed in the linear response regime during the experiments, we first measure the power dependence of their response and set the probe power of the following experiments to be ${\sim}\qty{10}{\decibel}$ below the power at which we observe the onset of nonlinearities in the form of a Duffing-like deformation of the absorption dips and/or a lineshape broadening due to nonlinear damping.
The first observation we make after junction writing is that the resonance frequencies are significantly shifted to lower values due to the additional Josephson inductance of the bJJs $L_\mathrm{J0} = \Phi_0/2\pi I_0$.
For cavity R1, this initial frequency shift from $\omega_\mathrm{c}$ (without bJJs) to $\omega_0$ (with bJJs) is as large as $\delta\omega_0=\omega_\mathrm{c}-\omega_0\approx2\pi\times \qty{175}{\mega\hertz}$ (see Fig.~\ref{fig:figure3}), for R2 and R3 the shifts are even larger, ${\sim}\qty{300}{\mega\hertz}$ and ${\sim}\qty{450}{\mega\hertz}$, respectively, cf.~Appendix\ref{app:add_data}.
In principle, we could already determine the Josephson inductances and critical currents at this point from the shift, but we cannot rule out the possibility that part of the shift has a different origin, such as e.g.~aging of the YBCO films or parasitic helium irradiation, both of which can increase the kinetic inductance of the cavity.
Hence, we determine $I_0$ from a more comprehensive dataset, which includes the bias flux dependence of~$\omega_0$.
To this end we vary the bias flux through the SQUIDs $\Phi_\mathrm{b}$ by sweeping a current $I_\mathrm{coil}$ through the attached coil and take a VNA trace of $S_{21}$ for each value of $I_\mathrm{coil}$.
From fits to the resulting resonance data, cf.~Appendix\ref{app:fitting} and Fig.~\ref{fig:figure3}\sublabel{a}, we obtain resonance frequency $\omega_0$ and linewidths $\kappa_0$ and $\kappa_\mathrm{ext}$ as a function of $\Phi_\mathrm{b}$.
Resonance frequency and total linewidth are periodically modulating with a period of one flux quantum $\Phi_0$ due to fluxoid quantization in the SQUID loop.
The modulation of $\omega_0$ is shown for R1 in Fig.~\ref{fig:figure3}\sublabel{b} and for R2 and R3 in Appendix\ref{app:add_data}.
It is non-hysteretic for all three cavities, and for R1 we find a (clearly detectable) modulation of the resonance frequency down to $\omega_0 = 2\pi\times \qty{4.39}{\giga\hertz}$ at $\Phi_\mathrm{b}/\Phi_0 \sim 0.48$, i.e., a tuning span of $\gtrsim \qty{310}{\mega\hertz}$.
For the even lower values around $\Phi_\mathrm{b}/\Phi_0 \approx 0.5$ the internal cavity linewidth gets too large to unambiguously identify $\omega_0$, but we estimate the total tuning range to be approximately $\qty{400}{\mega\hertz}$.
For the maximum flux responsivity, an important figure of merit for many applications, we find $\partial\omega_0/\partial\Phi_\mathrm{b} \approx 2\pi\times\qty{5}{\giga\hertz}\,\Phi_0^{-1}$ at $\Phi_\mathrm{b}/\Phi_0 \approx 0.48$.
R2 and R3 show even stronger modulations with total tuning ranges up to ${\gtrsim}\qty{1}{\giga\hertz}$ and flux responsivities up to $\partial\omega_0/\partial\Phi_\mathrm{b} \approx 2\pi\times\qty{24}{\giga\hertz}\,\Phi_0^{-1}$, cf.~Appendix\ref{app:add_data}, as one might have already assumed from the larger shift they experienced due to the junction writing.
With the resonance-frequency tuning-arcs, we have everything we need to extract more quantitative information about the bJJs and the SQUIDs.
For a symmetric SQUID (which we will assume for now) the relation between the externally induced flux $\Phi_\mathrm{b}$ and the total flux in the SQUID $\Phi$ (containing also flux from the circulating current in the loop) is given by
\begin{equation}
	\frac{\Phi}{\Phi_0} = \frac{\Phi_\mathrm{b}}{\Phi_0} - \frac{\beta_L}{2}\sin\left(\pi\frac{\Phi}{\Phi_0} \right)
\end{equation}
where $\beta_L = 2L_\mathrm{loop}I_0/\Phi_0 = L_\mathrm{loop}/\pi L_\mathrm{J0}$ is the SQUID screening parameter.
Solving this relation numerically and fitting
\begin{equation}
	\omega_0(\Phi_\mathrm{b}) = \frac{\tilde{\omega}_\mathrm{c}}{1 + \frac{L_\mathrm{J0}}{4\tilde{L}_\mathrm{r}\cos\left( \pi\frac{\Phi}{\Phi_0}\right)}}
	\label{eqn:fluxarc_prime}
\end{equation}
to the flux tuning data delivers as fit parameter the (mean) single-junction critical current $I_0$ or equivalently the resonance frequency of the cavity $\tilde{\omega}_\mathrm{c}$ before writing the bJJs.
Note that in the fit we keep $L_\mathrm{loop}$, the total cavity capacitance $C_\mathrm{tot}$ and the sweetspot frequency $\omega_{00} = \omega_0(\Phi = 0)$ fixed, and hence $I_0$ and $\tilde{\omega}_\mathrm{c}$ are not independent of each other.
The linear equivalent cavity inductance $\tilde{L}_\mathrm{r}$ is calculated via $\tilde{L}_\mathrm{r} = \pi^2/16 C_\mathrm{tot} \tilde{\omega}_\mathrm{c}^2$, cf.~Appendix\ref{app:LC_wJJ}.
As a result of the arc fit, we obtain the critical currents shown in Fig.~\ref{fig:figure3}\sublabel{c}, for R1 we get $I_0 = \qty{2.76}{\micro\ampere}$, for R2 and R3 we obtain $\qty{1.90}{\micro\ampere}$ and $\qty{1.51}{\micro\ampere}$, respectively.
These critical currents correspond to sweetspot Josephson inductances $L_{\mathrm{J}0}$ of $\qty{119}{\pico\henry}$, $\qty{173}{\pico\henry}$ and $\qty{219}{\pico\henry}$ for R1, R2, and R3, respectively.
In total, the $I_0$ and $L_{\mathrm{J}0}$ values vary only by roughly a factor of \num{2} between the three cavities, which seems smaller than what one would have expected from the scattering of the dc samples for identical doses.
Interesingly, they seem to vary even less within each SQUID, since the shape of the flux tuning arcs restricts possible asymmetries to $|\alpha_I| \lesssim 0.15$ for all three cavities~($\alpha_I$ is the critical-current asymmetry parameter, cf.~Appendix\ref{app:fluxSQUID}), which corresponds to ${\lesssim}\qty{30}{\percent}$ difference in the two critical currents.
Note that by post-annealing it would be possible to increase the critical currents in a controlled way, if desired~\cite{Karrer2024}.
The difference between $\omega_\mathrm{c}$ and $\tilde{\omega}_\mathrm{c}$ obtained from this approach is reasonably small, for all three resonances we get $\left(\omega_\mathrm{c} - \tilde{\omega}_\mathrm{c}\right)/\omega_\mathrm{c} \sim 0.02$, which we suspect to originate from aging of the YBCO film in the several months between the pre-characterization and the junction writing, and the corresponding increase of its kinetic inductance.
Further contributions might originate from the re-mounting and re-wirebonding, an underestimated loop inductance or weakly non-sinusoidal current-phase-relations.
Beyond the effects of the bJJs on the resonance frequency, we notice that their presence increases the cavity decay rates and that the junction-induced loss rate grows with decreasing junction critical current, cf.~Fig.~\ref{fig:figure3}\sublabel{d}.
To isolate the additional losses introduced by junction writing, we introduce the junction loss factor at the flux sweetspot
\begin{equation}
	\frac{1}{Q_\mathrm{J0}} = \frac{1}{Q_{\mathrm{int}0}} - \frac{1}{Q_\mathrm{int, c}},
	\label{eqn:lossJ0_def}
\end{equation}
where $1/Q_{\mathrm{int}0} = \kappa_{\mathrm{int}0}/\omega_{00}$ with the internal linewidth at the flux sweetspot $\kappa_{\mathrm{int}0}$ and $1/Q_\mathrm{int, c} = \kappa_\mathrm{int, c}/\omega_\mathrm{c}$ before introducing the bJJs.
As a result, we find junction-induced loss factors in the $10^{-3}$ range and a considerable dependence of $1/Q_{\mathrm{J} 0}$ on the critical current, which seems to be following approximately ${\propto}I_0^{-2}$.
When adding bias flux and tuning the resonance frequency, the junction-induced losses increase further, cf.~Fig.~\ref{fig:figure3}\sublabel{a}, likely due to the flux-induced reduction of the critical currents, but potentially also by additional contributions from internal and external flux noise.
A more quantitative discussion of the sweetspot loss factors and a simple model will be presented below.
The observed $1/Q_{\mathrm{J}0}$ are equivalent to bJJ-related quality factors between ${\sim}200$~(R3) and ${\sim}620$~(R1), which seems not overly impressive when considering how high the quality factors of superconducting circuits can get.
However, there are several important things to consider.
First, we believe that these low quality factors are partly caused by the high measurement temperature and thus they might be unavoidable for $\unit{\micro\ampere}$-$I_0$ SQUID cavities at elevated temperatures, cf.~also~\cite{Uhl2023, Uhl2024, Uhl2024a}.
In that case, a possible solution for some applications might be a re-design using much smaller SQUID loops, larger critical currents and lower linear circuit inductances to get cavities with similar characteristics, but higher~$Q$.
Secondly, despite these rather low quality factors, the circuits presented here might still be the best option to date for highly tunable microwave circuits at liquid helium temperatures or perspectively in tesla magnetic fields, since e.g.~aluminum circuits are no longer superconducting at all under these conditions.
And lastly, for some applications large loss factors such as the ones reported here are not harmful or even desired, such as parametric amplifiers and dispersive SQUID sensors with maximized $\kappa_\mathrm{ext}$ or dissipative radiation-pressure systems~\cite{Kazouini2026}, respectively.
\vspace{-3mm}
\subsection*{The single-junction Fraunhofer tuning}
\vspace{-2mm}
\begin{figure*}
	\includegraphics{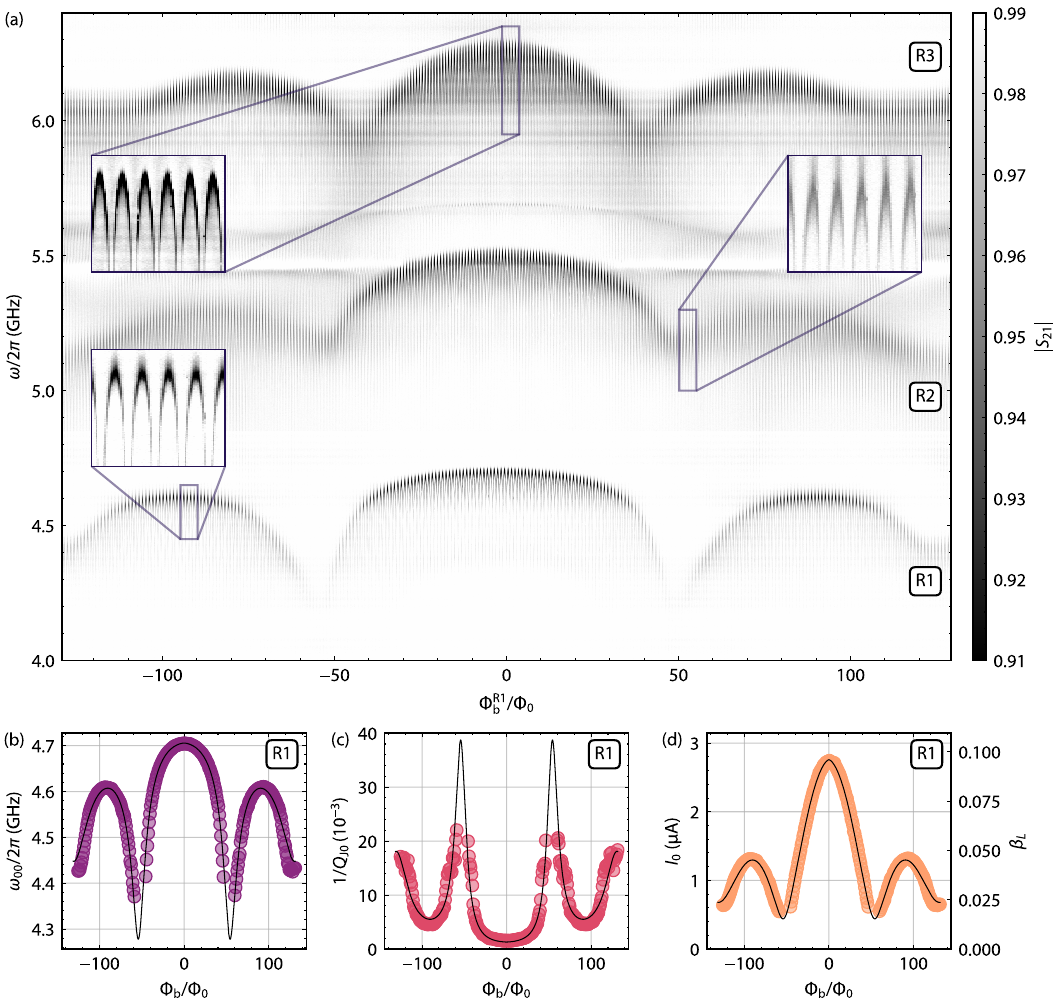}
	\vspace{-3mm}
	\titlecaption{Single-bJJ Fraunhofer-diffraction decouples resonance frequency from flux responsivity and allows modeling losses.}{%
		\sublabel{a}~Gray-scale-coded transmission $|S_{21}|$ vs VNA probe frequency (covering the resonance frequencies of all three cavities) and vs bias flux coupling into the SQUID of R1. Beyond a few flux quanta $\Phi_0$, all three modes show a Fraunhofer-like modulation of their resonance frequencies enveloping the periodic modulations of the SQUIDs, which are still clearly visible over the complete field range, even around the Fraunhofer minima or at the first side-maxima, cf.~the three zoom windows. Note here that the zoom window for R2 reveals skewed flux arcs, a signature for Josephson diodes as discussed in detail in Appendix~\ref{app:add_data}. Slightly above $\qty{5.5}{\giga\hertz}$ a weakly modulating cavity from a second on-chip feedline can be seen (bJJs with a lower ion dose, so less modulation), that appears due to a small amount of parasitic cross-talk between the feedlines. The sweetspots of the individual SQUID flux arcs cover a frequency range of several hundred~\unit{\mega\hertz}.
		\sublabel{b}~Sweetspot resonance frequency $\omega_{00}$ of R1 vs $\Phi_\mathrm{b}/\Phi_0$. We define the sweetspot as the point of maximum $\omega_0$ and vanishing $\partial\omega_0/\partial\Phi_\mathrm{b}$ of each SQUID arc. Although in \sublabel{a} it is apparent that the modulation goes lower than to $\qty{4.35}{\giga\hertz}$, it is difficult to extract reliable values there due to the shallowness of the cavity resonances. This sweetspot modulation allows tuning the resonance frequency without increasing the sensitivity to flux noise (to first order) within a tuning range of several hundred~\unit{\mega\hertz}.
		\sublabel{c}~Junction-induced sweetspot loss factor $1/Q_{\mathrm{J}0}$, showing an inverse Fraunhofer-like modulation.
		\sublabel{d}~Critical junction current $I_0$ and screening parameter $\beta_L$ as a function of $\Phi_\mathrm{b}/\Phi_0$. The large modulation of $I_0$ from $I_0^\mathrm{max} = \qty{2.76}{\micro\ampere}$ down to $I_0^\mathrm{min} \sim \qty{500}{\nano\ampere}$ indicates that the two junctions in the SQUID have closely matching parameters and a high-quality junction barrier. Symbols in \sublabel{b}\textendash\sublabel{d} are data. Line in \sublabel{d} is a phenomenological fit curve, cf.~Appendix\ref{app:Fraunfit}, based on which we calculate $\omega_{00}$~[line in \sublabel{b}] and fit the loss-factor [line in~\sublabel{c}] using a single fit parameter, cf.~main text.
	}
	\label{fig:figure4}
\end{figure*}
In the data discussed in Fig.~\ref{fig:figure3}, we applied only very small out-of-plane magnetic fields to the samples.
The effective area of the SQUID loops without the ground planes, estimated from numerical simulations using \textit{3D-MLSI}~\cite{Khapaev2001}, is around $A_\mathrm{eff} \approx \qty{200}{\micro\meter\squared}$ and hence to introduce one flux quantum into the SQUID, the mean local field (which is enhanced compared to the applied field due to flux focusing effects of the ground planes) is on the order of $B_{0} = \Phi_0/A_\mathrm{eff} \sim \qty{10}{\micro\tesla}$.
On the other hand, we have observed in Fig.~\ref{fig:figure2}\sublabel{b} that an out-of-plane field on the order of ${\sim}\qty{0.55}{\milli\tesla}$ will suppress the critical current of a single bJJ due to the Fraunhofer-like diffraction pattern.
So a natural question is:~Will we also observe the fingerprints of such a Fraunhofer modulation in the microwave response if we increase the external field significantly beyond the few flux quanta of Fig.~\ref{fig:figure3}?
Figure~\ref{fig:figure4} reveals the answer:~We do indeed.
Superimposed on the periodic SQUID modulations of the absorption resonances, we find a modulation-envelope with a first minimum at approximately $\pm 53 \Phi_0$, which corresponds to roughly $\qty{0.53}{\milli\tesla}$, cf.~Fig~\ref{fig:figure4}\sublabel{a}.
For even larger fluxes, the sweetspot frequencies bounce back from the minima and symmetrically form a second dome on both sides of the central dome, just with lower maxima and narrower width compared to the main peak, cf.~Fig~\ref{fig:figure4}\sublabel{b}.
The interference superposition is completely analogous to that of a double-slit experiment with finite single-slit diffraction in optics.
Assuming that at the sweetspots of all the flux arcs $\Phi_\mathrm{b} = n\Phi_0, n\in \mathbb{Z}$, and hence that there is no screening current in the SQUID loop, we can determine the corresponding (mean) critical junction current as a function of the external field as shown in Fig.~\ref{fig:figure4}\sublabel{d} for R1 and in Appendix\ref{app:add_data} for R3.
An exemplary theoretical treatment of a cavity with Fraunhofer-tunable JJs is given in Appendix\ref{app:fluxFraun}.
The shape in Fig.~\ref{fig:figure4}\sublabel{d} is remarkably similar to the one found in dc, ~Fig.~\ref{fig:figure2}\sublabel{b} again, just lower by a factor of ${\sim}3.5$ and symmetrized due to the microwave experiment probing negative and positive critical current simultaneously.
Such a combination of SQUID oscillations and single-bJJ field-tuning -- which is faciliated by the orientation of the junction barriers perpendicular to the film plane -- might be a very useful resource, since it allows to decouple the SQUID cavity resonance frequency from its flux responsivity~$\partial\omega_0/\partial\Phi_\mathrm{b}$.
As one example this is interesting for applications which require tuning $\omega_0$ but preferably without increasing its sensitivity to magnetic field or flux noise, which can be a problem particularly in large-SQUID circuits~\cite{Paradkar2025}.
The collection of sweetspot frequencies for cavity R1 here spreads over several hundred \unit{\mega\hertz}, cf.~Fig.~\ref{fig:figure4}\sublabel{b}, which is roughly the same tuning range one can achieve with the SQUID-tuning over one flux quantum, cf.~Fig.~\ref{fig:figure3}\sublabel{b}.
At all the flux sweetspots, however, the flux responsivity is zero and the cavity is (to first order) insensitive to field noise (as long as the noise-field is homogeneous across SQUID and junctions).
Similarly, one gains the freedom to choose the resonance frequency for any desired finite flux responsivity in e.g.~optomechanical or photon-pressure systems with SQUID circuits~\cite{Rodrigues2019, Zoepfl2020, Schmidt2020, Schmidt2024, Bothner2021, Kazouini2026}.
Beyond aspects of the flux responsivity, the combination of both tuning mechanisms together can increase the overall tuning range, e.g.~in cases, where the SQUID screening parameter is large, and the absence of a strict periodicity could be interesting for absolute field- and flux-detection~\cite{Guenzler2021}.
Furthermore, if one chooses identical zero-field critical currents for the junctions, but different widths $a$, it should be possible to adjust the SQUID asymmetry using the external field, since the widths of the Fraunhofer-tuning curves change with $a$.
Finally, such a single-junction Fraunhofer tuning could open a new degree of freedom for JJ-array applications such as heavy fluxonium qubits or parametric amplifiers without resorting to the larger complexity of SQUID arrays.
We not only observe the Fraunhofer-like modulation in $\omega_{00}$, but also in the loss rate $1/Q_{\mathrm{J}0}$, cf.~Fig.~\ref{fig:figure4}\sublabel{c}.
The losses behave opposite to the critical current and the resonance frequency, they increase for decreasing $I_0$ and $\omega_{00}$, and have their minima at the maxima of critical current and resonance frequency.
Based on the earlier observed trend $1/Q_{\mathrm{J}0} \propto I_0^{-2}$ this is also not too surprising.
Due to the amount of data, we can now compare a simple but quantitative model for the losses to the experimental result.
If we treat each bJJ as a parallel combination of a Josephson inductance and a field-independent resistor $R_\mathrm{J}$, basic circuit modeling predicts
\begin{equation}\label{eqn:losses_main}
	\frac{1}{Q_{\mathrm{J}0}} = \frac{\omega_{00}L_{\mathrm{J}0}^2}{R_\mathrm{J}\left(2L_\mathrm{r} + L_\mathrm{J0} \right)},
\end{equation}
a derivation can be found in Appendix\ref{app:losses}.
Sweetspot frequency and sweetspot critical current $I_0 = \Phi_0/2\pi L_{\mathrm{J}0}$ are both functions of the external magnetic field and so in order to fit Eq.~(\ref{eqn:losses_main}) to the loss factor data with a single parameter $R_\mathrm{J}$, we first model the sweetspot critical currents $I_0(\Phi_\mathrm{b}/\Phi_0)$ using a phenomenological equation, cf.~Appendix\ref{app:Fraunfit} and Fig.~\ref{fig:figure4}\sublabel{d} for the result.
Using this analytical function for $I_0$ as well as $L_\mathrm{r} = \tilde{L}_\mathrm{r}$, we calculate $L_{\mathrm{J}0}$ and $\omega_{00}$, cf.~Fig.~\ref{fig:figure4}\sublabel{b}, and finally fit $1/Q_{\mathrm{J}0}$.
We obtain very good agreement between the model and the data, cf.~Fig.~\ref{fig:figure4}\sublabel{c}, and as fit parameter value we obtain $R_\mathrm{J} = \qty{101}{\ohm}$, a result that matches perfectly the trend from the resistance evaluation of the dc bJJs, cf.~Appendix\ref{app:add_data}.
Note that for $L_\mathrm{r} \gg L_{\mathrm{J}0}/2$ and constant $\omega_{00}$, $L_\mathrm{r}$ and $R_\mathrm{J}$, Eq.~(\ref{eqn:losses_main}) is actually equivalent to $1/Q_{\mathrm{J}0} \propto I_0^{-2}$.
The model and the agreement of $R_\mathrm{J}$ with the results from the dc IVC measurements, cf.~Appendix\ref{app:add_data} again, suggests that the resistance entering the equations for the microwave losses $R_\mathrm{J}$ is identical to the junction resistance in the voltage state $R_\mathrm{n}$, an observation in contrast to low-$T_\mathrm{c}$ tunnel junctions and potentially related to the incompletely gapped d-wave order parameter in YBCO.
However, the model also tells us how to potentially achieve high-$Q_{\mathrm{J}}$ YBCO bJJ circuits.
For fixed $\omega_{00}$ one would need either higher junction resistances, which might be possible with lower temperatures, higher resistivity YBCO films or different junction geometries such as constrictions~\cite{Schmid2025}, or larger linear inductance contributions $L_\mathrm{r}$ and higher critical currents~$I_0$.
\vspace{-3mm}
\subsection*{Temperature-dependence}
\vspace{-2mm}
\begin{figure*}
	\includegraphics{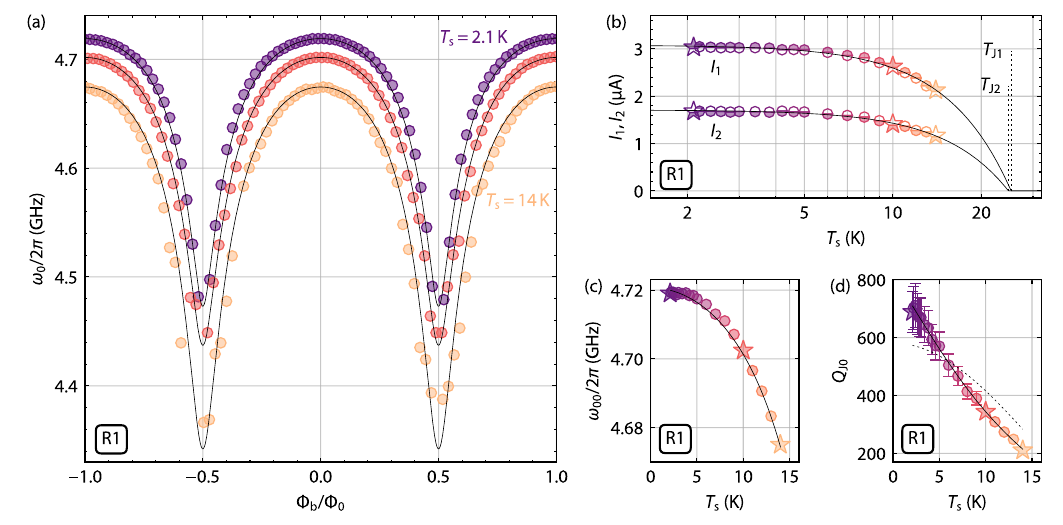}
	\vspace{-4mm}
	\titlecaption{Operating flux-tunable YBCO SQUID cavities between $\qty{2.1}{\kelvin}$ and $\qty{14}{\kelvin}$.}{%
		\sublabel{a}~Flux-tuning arcs of the resonance frequency $\omega_0$ of R1 at three different sample temperatures $T_\mathrm{s} = \qty{2.1}{\kelvin}$, $\qty{10}{\kelvin}$ and $\qty{14}{\kelvin}$, plotted in purple, red, and orange, respectively. With increasing temperature, the sweetspot resonance frequency decreases, and the tuning range increases from $\omega_{00} - \omega_\mathrm{min} \sim 2\pi\times\qty{240}{\mega\hertz}$ at~$\qty{2.1}{\kelvin}$ to ${\sim}2\pi\times\qty{320}{\mega\hertz}$ at~$\qty{14}{\kelvin}$. Circles are experimental data, lines are fit curves considering potentially different critical currents of the two bJJs in the SQUID, i.e., an asymmetric SQUID.
		\sublabel{b}~Critical currents $I_1, I_2$ of the two bJJs in the SQUID vs $T_\mathrm{s}$ obtained from the arc fits in panel~\sublabel{a}. Top-filled symbols correspond to $I_1$, bottom-filled symbols to $I_2$. Over the complete experimental $T_\mathrm{s}$-range, the two critical currents differ by a factor of ${\sim}1.8$, which corresponds to an asymmetry parameter $\alpha_I \sim 0.28$. Symbols are data, lines are fit curves, cf.~main text. The effective transition temperatures of both bJJs is $T_{\mathrm{J}1} \approx T_{\mathrm{J}2} \approx \qty{25}{\kelvin}$. The zero-temperature critical currents are $I_{10} = \qty{3.08}{\micro\ampere}$ and $I_{20} = \qty{1.70}{\micro\ampere}$.
		\sublabel{c}~Sweetspot resonance frequency $\omega_{00}$ vs sample temperature $T_\mathrm{s}$. With increasing $T_\mathrm{s}$ the sweetspot frequency decreases due to the temperature-dependent bJJ critical currents. Total change of $\omega_{00}$ over the investigated temperature range is almost one order of magnitude smaller than the tuning range via $\Phi_\mathrm{b}$. Symbols are data, line is a theory curve based on the fit curves of $I_1, I_2$ in \sublabel{b}.
		\sublabel{d}~Junction-induced sweetspot quality factor $Q_{\mathrm{J}0}$ vs sample temperature $T_\mathrm{s}$. With increasing temperature the quality factor decreases by roughly a factor $3.5$ from ${\sim}700$ at the lowest temperature to ${\sim}200$ at the highest one. Symbols are data, dashed line is a theory curve based on the fits for $I_1, I_2$ with a temperature-independent junction resistance $R_\mathrm{J}$, solid line is a fit curve with a temperature-dependent $R_\mathrm{J}$, for details see main text. For a discussion of the error bars, cf.~Appendix\ref{app:errors}; when the error is smaller than the symbol size, the error bars are omitted.
		Star symbols in panels~\sublabel{b}{\textendash}\sublabel{d} correspond to the datasets shown in panel~\sublabel{a}.
	}
	\vspace{-0mm}
	\label{fig:figure5}
\end{figure*}
As a last experiment, we investigate the temperature-dependence of the SQUID cavities in a temperature range $\qty{2.1}{\kelvin} \leq T_\mathrm{s} \leq \qty{14}{\kelvin}$.
This not only allows us to investigate the bJJ characteristics as a function of temperature, but also provides some insight into the possible operation range of the devices with respect to higher-temperature applications, e.g.~the spin-resonance study of phase transitions in magnetic materials~\cite{Miksch2021}.
Since variation of the sample temperature is not feasible in liquid helium, we mount the device into the evacuable microwave insert of a liquid helium cryostat.
The insert is similarly equipped with dc cables, microwave cables, attenuators and a HEMT as the dipstick used for the experiments above, but in addition contains a temperature sensor and a heating resistor for feedback-stabilized control of $T_\mathrm{s}$ with fluctuations on the sub-\unit{\milli\kelvin} level, for details see Appendix\ref{app:setups}.
By vacuum pumping on the helium reservoir of the cryostat and a small amount of helium exchange gas in the vacuum compartment with the sample, we reach down to $T_\mathrm{s}^\mathrm{min} \approx\qty{2}{\kelvin}$, mainly limited by the heat load of the HEMT operation.
In the experiment, we first stabilize the temperature to a desired value, and then perform a bias flux sweep as described in the context of Fig.~\ref{fig:figure3}.
Figure~\ref{fig:figure5} shows the most important insights we gained from analyzing the flux- and temperature-dependence of the spectroscopically determined cavity characteristics.
We can clearly detect flux-tuning curves up to $\qty{14}{\kelvin}$, above which the internal linewidth gets too large to analyze the resonances for the complete range of $\Phi_\mathrm{b}$, and hence we limit the analysis to $T_\mathrm{s}\leq\qty{14}{\kelvin}$.
The overall shape of the tuning arcs is only weakly dependent on temperature, but they shift slightly downwards in frequency and the tuning range slightly increases with increasing $T_\mathrm{s}$.
However, the tuning range is reduced compared to the dataset shown in Fig.~\ref{fig:figure3}\sublabel{b}.
Now, it is only on the order of ${\sim}\qty{240}{\mega\hertz}$ for the lower temperatures and ${\sim}\qty{320}{\mega\hertz}$ for the higher temperatures.
In addition to the reduced tuning range, we observe inverted sweetspots at $\pm0.5\Phi_0$, which indicates an asymmetric SQUID.
We believe this newly observed SQUID asymmetry originates from variations in $I_0$ between different cooldowns and over time, and not from the different setup or the absence of liquid helium.
To model the arcs, we implement the theory for an asymmetric SQUID circuit, in which the two junctions have the critical currents $I_1 = I_0(1 + \alpha_I)$ and $I_2 = I_0(1 - \alpha_I)$ with the asymmetry parameter $\alpha_I$, cf.~also Appendix\ref{app:fluxSQUID}.
The flux equation becomes
\begin{equation}
	\frac{\Phi}{\Phi_0} = \frac{\Phi_\mathrm{b}}{\Phi_0} - \frac{\beta_L}{2}\frac{\left(1 - \alpha_I^2 \right)}{\sqrt{1 + \alpha_I^2 \tan^2\left(\pi\frac{\Phi}{\Phi_0} \right)}}\sin\left(\pi\frac{\Phi}{\Phi_0} \right),
\end{equation}
i.e., the critical-current asymmetry modifies the relation between the external flux and the total flux in the SQUID, and the resonance frequency as a function of the total flux $-0.5\leq \Phi/\Phi_0 \leq 0.5$ follows
\begin{equation}
	\omega_0(\Phi_\mathrm{b}) = \frac{\tilde{\omega}_\mathrm{c}^*}{1 + \frac{L_{\mathrm{J}0}}{4\tilde{L}_\mathrm{r}\cos\left( \pi\frac{\Phi}{\Phi_0}\right)\sqrt{1 + \alpha_I^2\tan^2\left( \pi\frac{\Phi}{\Phi_0}\right)}}}.
	\label{eqn:w0_asym}
\end{equation}
By fitting the data for each temperature individually with $\alpha_I$ and $I_0$ as fit parameters, $\tilde{L}_\mathrm{r}$ fixed from the results at $\qty{4.2}{\kelvin}$, and $\tilde{\omega}_\mathrm{c}^*$ constant for all $T_\mathrm{s}$, we extract the critical junction currents $I_1, I_2$ as a function of temperature, cf.~Fig.~\ref{fig:figure5}\sublabel{b}.
The resonance frequency $\tilde{\omega}_\mathrm{c}^* = 2\pi\times \qty{4.825}{\giga\hertz}$ is somewhat larger than $\tilde{\omega}_\mathrm{c}$, likely due to the low-pressure helium-gas environment compared to liquid helium, which slightly reduces the effective CPW permittivity and hence $C_\mathrm{tot}$.
For a discussion on the potential temperature-dependence of $\tilde{\omega}_\mathrm{c}^*$, $\tilde{L}_\mathrm{r}$ and $L_\mathrm{loop}$, which we neglect here, see~Appendix\ref{app:dev_parameters}.
The critical currents at the lowest temperatures are $I_1\approx \qty{3.03}{\micro\ampere}$ and $I_2\approx \qty{1.68}{\micro\ampere}$, so $I_0 = (I_1 + I_2)/2 \approx \qty{2.36}{\micro\ampere}$ is somewhat reduced compared to the earlier $\qty{4.2}{\kelvin}$ data.
From the decreased $I_0$ and the increased asymmetry, we suspect that (for an unknown reason) $I_2$ is smaller than it was in the earlier cooldown in liquid helium and $I_1$ remained approximately constant.
Both critical currents show a smooth downwards trend with temperature up to $\qty{14}{\kelvin}$, where they reach $I_1 \approx \qty{2.13}{\micro\ampere}$ and $I_2 \approx \qty{1.17}{\micro\ampere}$, and so $\alpha_I \approx 0.29$ for the complete temperature range with a slight tendency to increase with $T_\mathrm{s}$.
The temperature dependence of both critical currents can be well described by
\begin{equation}
	I_{k}(T_\mathrm{s}) = I_{k0}\left[1 - \left( \frac{T_\mathrm{s}}{T_{\mathrm{J}k}}\right)^2\right], ~~~k \in \{1, 2\}
	\label{eqn:NotBardeen}
\end{equation}
with $I_{0k}$ the critical current of the $k$-th junction at $T_\mathrm{s} = 0$, and $T_{\mathrm{J}k}$ the effective transition temperature of the corresponding bJJ, i.e., the temperature at which $I_k = 0$.
The actual temperature dependence at $T_\mathrm{s} > \qty{14}{\kelvin}$ and $T_{\mathrm{J}k}$ could also be different from Eq.~(\ref{eqn:NotBardeen}), but since there is not a universal law for the $T_\mathrm{s}$-dependence of $I_0$ in YBCO JJs or d-wave superconductors, and considering the currently limited amount of data, we are content with a general function that describes the dataset for $T_\mathrm{s} \leq \qty{14}{\kelvin}$ sufficiently well.
Finally, we consider the junction-related sweetspot quality factor $Q_\mathrm{J0}$ as a function of temperature, cf.~Fig.~\ref{fig:figure5}\sublabel{d}.
The experimental values decrease from ${\sim}700$ at $\qty{2.1}{\kelvin}$ to ${\sim}200$ at $\qty{14}{\kelvin}$, and the $Q_{\mathrm{J}0}$ curve seems not to be described anymore by Eq.~(\ref{eqn:losses_main}) with $R_\mathrm{J} = \qty{101}{\ohm}$, which is added to the figure as dashed line, or any other temperature-independent $R_\mathrm{J}$.
While the experimental curve has a positive curvature, the theory curve with a temperature-independent $R_\mathrm{J}$ clearly has a negative curvature, saturates hand in hand with $I_1$ and $I_2$ for $T_\mathrm{s} \rightarrow 0$, and varies by much less than the experimental data over the complete $T_\mathrm{s}$ range.
However, if we fit the model with a temperature-dependent resistance $R_\mathrm{J}(T_\mathrm{s}) = R_0 + r_0(T_\mathrm{s} - T_\mathrm{ch})^2$ using $R_0, r_0$, and $T_\mathrm{ch}$ as fit parameters, we find excellent agreeement again, cf.~solid line in Fig.~\ref{fig:figure5}\sublabel{d}.
The resulting $R_\mathrm{J}(T_\mathrm{s})$ is between $\qty{127}{\ohm}$ for the lowest temperatures and $\qty{75}{\ohm}$ for the highest ones, and both the general trend of the decreasing resistance with increasing temperature and the overall shape of $R_\mathrm{J}(T_\mathrm{s})$, cf.~Appendix\ref{app:add_data}, is in good agreement with other results from dc experiments on high-dose helium-ion bJJs in YBCO~\cite{Cybart2015, Couedo2020}.
The good agreement between model and data allows us to make a prediction for the achievable quality factors in the \unit{\milli\kelvin} regime.
The cavity discussed in Fig.~\ref{fig:figure5} seems to be limited to $Q_{\mathrm{J}0}\lesssim 1000$, unless some unexpected steep upwards trend sets in below $\qty{2}{\kelvin}$, something we intend to explore in future studies.
A redesign of the cavity and the bJJs, however, could achieve comparable $\omega_{00}$ and comparable frequency tuning ranges to R1, but with an up to tenfold increased quality factor $Q_\mathrm{J0}$.
One possible choice would be to reduce $L_\mathrm{r}$, $L_\mathrm{loop}$ and $L_{\mathrm{J}0}$ by a factor of ${\sim}3$, and to simultaneously increase the junction resistance $R_\mathrm{J}$ by a factor of ${\sim}3$.
The former is rather straightforward to accomplish with the only drawback of smaller SQUID loops, the latter might be achieved using higher-resistivity YBCO films, by in-plane rotation of the bJJ angle with respect to the YBCO symmetry axes, exploiting the conductivity anisotropy of YBCO~\cite{LeFebvre2023}, or by junction-geometry adjustments such as nano-constrictions~\cite{Schmid2025}.
Then, with $Q_{\mathrm{J}0}\sim 10^4$, the resulting SQUID cavities would be able to compete with existing low-$T_\mathrm{c}$ SQUID devices in terms of their losses~\cite{PalaciosLaloy2008, Pogorzalek2017, Bothner2022, Paradkar2025}.
\vspace{-3mm}
\section*{Discussion}
\vspace{-2mm}
In this work, we investigated magnetic-field-tunable coplanar waveguide cavities, based on the high-$T_\mathrm{c}$ cuprate superconductor YBCO, with integrated helium-ion-written Josephson junctions.
We observed periodic SQUID oscillations of the resonance frequencies up to the \unit{GHz} range with a superimposed modulation due to the Fraunhofer-like diffraction tuning of the individual junctions.
The unusual Fraunhofer modulation effect originating from the out-of-plane junction barriers not only suggests high-quality JJs, but also provides exciting possibilities, such as decoupling resonance frequency from flux responsivity~(i.e.~minimized noise sensitivity) or in-situ tunable SQUID asymmetry.
Furthermore, we demonstrated that the cavities can be operated with a nearly constant frequency-tunability in a large temperature range up to $\qty{14}{\kelvin}$, more than an order of magnitude higher than what is possible with the standard material aluminum.
These results, in combination with the enormous flexibility of helium-ion-based Josephson junctions, highlight the great potential of the investigated devices for various high-temperature and perspectively high-field applications, such as SQUID optomechanics, photon-pressure systems, space missions, electron-spin-resonance and phase-transition spectrometers, high-field parametric amplifiers, dispersive SQUID microscopy, or hybrid systems with magnons.
To the best of our knowledge and despite the moderate absolute values, the cavities presented in this work constitute the highest-quality-factor YBCO SQUID cavities reported to date.
Based on a simple model, we have analyzed the junction-related energy losses of the cavities, which will likely enable lower losses in future devices, potentially to a level where YBCO SQUID circuits operated at \unit{\milli\kelvin} will show loss levels comparable to similar SQUID devices made from other materials.
If the losses can be made sufficiently small, this work might even pave the way for implementing YBCO transmon or fluxonium qubits for sensing applications, quantum hybrid systems in large magnetic (in-plane) fields or to study some yet elusive aspects of helium-ion bJJs such as critical-current or quasiparticle noise.
Other interesting questions to consider in follow-up experiments are the cavity characteristics in large magnetic in-plane fields, the cavity and junction properties at \unit{\milli\kelvin} temperatures, and the impact of junction width, junction angle, junction-inductance participation ratio or junction-type~(e.g.~usual bJJ vs nano-constriction) on device performance.
\vspace{-4mm}
\begin{acknowledgments}
	\vspace{-2mm}
	The authors warmly thank Markus Turad, Ronny Löffler (instrument scientists of the core facility \lisaplus), and Christoph Back for technical support.
	This research received funding from the Vector Stiftung via project number \mbox{P2023-0201} and from the Deutsche Forschungsgemeinschaft (DFG) via grant numbers 490939971 (BO~\mbox{6068/1-1}) and 511315638 (BO~\mbox{6068/2-1}). M.K.~gratefully acknowledges financial support by the Studienstiftung des deutschen Volkes.
	We also gratefully acknowledge support by the COST actions NANOCOHYBRI (CA16218) and SUPERQUMAP (CA21144).
\end{acknowledgments}
\vspace{-3mm}
\subsection*{Data availability}
\vspace{-2mm}
All data presented in this paper, including raw data and processing code, will be publicly available on the repository Zenodo upon peer-reviewed publication of this work.
\vspace{-3mm}
\subsection*{Competing interests}
\vspace{-2mm}
The authors declare no competing interests.
\vspace{-3mm}
\subsection*{Author contributions}
\vspace{-2mm}
K.F.~designed and fabricated the devices, conducted the experiments, analyzed the data, prepared the figures and contributed to the first draft of the manuscript.
T.J.G.-M.~contributed to the experiments, analyzed the data, and prepared the figures.
B.W.~contributed to the experiments and to the measurement code.
M.K.~and C.S.~contributed to sample fabrication.
R.K.~and D.K.~contributed to project funding and participated in scientific discussions.
D.B.~conceived the experiment, acquired the project funding, supervised all aspects of the study and wrote the first draft of the manuscript.
All authors discussed the results and contributed to manuscript revisions.

\clearpage
\setcounter{subsection}{0}
\section*{Appendix}
\label{sec:appendix}
\makeatletter
\renewcommand{\@seccntformat}[1]{%
	Appendix~\csname the#1\endcsname:
}
\makeatother
\subsection{Sample fabrication and preparation}
\label{app:devicefab}
\vspace{-2mm}
We describe the fabrication and preparation procedure of both dc and microwave chips combined here, although some steps differ between the two chips.
When none of the two chips is mentioned explicitly, the corresponding step is performed for both chips identically.

\begin{itemize}[leftmargin = 0.0pt, label = {}]

	\item \textit{Step 1 -- Wafer dicing:}\\
	The first step of the sample fabrication is dicing of a $2\,$inch wafer into individual $\qty{10}{\milli\meter}\times\qty{10}{\milli\meter}$ large chips for further processing.
	The wafer is a $\qty{500}{\micro\meter}$ thick MgO substrate, covered with a bilayer of $\qty{50}{\nano\meter}$ M-type YBCO and $\qty{20}{\nano\meter}$ gold by the company Ceraco.
	The YBCO film is deposited using reactive co-evaporation and the gold is deposited \textit{in-situ} on top of the YBCO by evaporation.
	Before dicing, the complete wafer is covered with a layer of photoresist for protection from water and dirt.
	After dicing, the individual chips are cleaned by soaking them in multiple subsequent ultrasonic baths of acetone and isopropanol directly before step~2.

	\item \textit{Step 2 -- Bilayer patterning with lithography and Ar ion milling:}\\
	As a next step, the individual chips are covered with a $\qty{\sim400}{\nano\meter}$ thick layer of positive photoresist (ma-P~1205) by spin-coating, and the desired dc/microwave patterns are transferred to the resist by maskless optical lithography ($\lambda_\mathrm{Lit} = \qty{365}{\nano\meter}$) and resist development in the developer ma-D~331/S for $\qty{30}{\second}$.
	Afterwards, the chips are mounted into a high-vacuum ion-milling chamber and the unprotected film parts of both gold and YBCO are removed in a single continuous process by Ar ion milling.
	The remaining resist is washed off by multiple subsequent ultrasonic baths of acetone, followed by multiple ultrasonic baths of isopropanol.

	\item \textit{Step 3 -- Patterning and wet removal of gold:}\\
	The third step starts identically to step~2 with spin-coating of photoresist and optical lithography, but this time a different pattern is transferred to the chip with the goal of removing the gold from the YBCO at all places where it is undesired.
	For the microwave chip, this concerns most of the chip surface except for a $\qty{\sim 1}{\milli\meter}$ wide stripe at the chip edges, which is used for wirebonding, and small structures in the middle of the chip, which are later used for navigation and focusing of the He ion beam.
	On the dc chip, the gold is removed from the bridge structures for the later bJJs and in between the contact pads for current and voltage leads.
	After lithography and resist development (cf.~step~2), the pattern is transferred into the gold film by wet-etching in a $50/50$ bath of TechniEtch$^\mathrm{TM}$~ACI2 and water for $\qty{40}{\second}$.
	Finally, cleaning by soaking in ultrasonic baths of acetone and isopropanol is performed, as in steps~1 and~2.

	\item \textit{Step 4 -- Pre-characterization (microwave chip only):}\\
	At this point, we pre-characterize the microwave resonators to obtain their resonance frequencies and quality factors before introducing the He-bJJs.
	Hence, the microwave chip is wirebonded into a suitable microwave PCB and both are mounted into a radiation-tight copper housing.
	The wirebonds connect the center conductor of the PCB CPW to the center conductor of the on-chip CPW, the PCB ground plane to the on-chip ground plane all around the chip, and on-chip ground plane to on-chip ground plane across the on-chip CPW~(air bridges).
	Then, the microwave box is attached to the end of a microwave-compatible dipstick for measurements in liquid helium transport dewars, immersed in liquid helium and characterized by microwave spectroscopy using a VNA.
	After this pre-characterization, the chip is demounted from the copper box and PCB again (airbridge wirebonds are not removed), and prepared for helium ion irradiation.

	\item \textit{Step 5 -- Josephson junction patterning with He$^+$ ions:}\\
	The chips are wirebonded to an aluminum stub and loaded into a helium ion microscope~(HIM) for introduction of the Josephson junctions by focused-ion-beam irradiation.
	Gold markers on the chips are used to navigate to a position close to the structures destined for JJ irradiation without illuminating the critical YBCO parts, cf.~inset of Fig.~\ref{fig:figure2}\sublabel{a}, and the gold markers closest to the future JJs are used to focus the ion beam for each irradiation to keep the spot size as small as possible.
	Then, for each JJ, a single straight line with the desired ion line dose $D_\mathrm{\ell}$ and an ion energy $E_\mathrm{ion} = \qty{30}{\kilo\electronvolt}$ is irradiated across the YBCO bridges.
	Irradiation of the dc chip was done in two independent HIM sessions several weeks apart and with measurements of the first batch of 18 bJJs in between; irradiation of all bJJs on the microwave chip has been done in a single HIM session several months after the second dc chip irradiation.

	\item \textit{Step 6 -- Mounting and wirebonding:}\\
	After ion irradiation, the chips are removed from the ion microscope and mounted to an appropriate PCB (different layouts for microwave and dc), where they are wirebonded for the upcoming experiments.
	For the dc experiments, this is done in batches of six bJJs per cooldown.
	We keep the time spans between ion irradiation and the first cooldown and in between different cooldowns as short as possible to minimize the drift of critical currents over time, which likely appears due to thermal diffusion of displaced oxygen atoms near the JJ barriers~\cite{Karrer2024}.
	For the microwave chip, we proceed completely analogously to step~4.
	Due to unexpected delays, however, approximately ten weeks passed between the bJJ writing and the microwave measurements presented in this work.
	In all experiments with bJJs, both sample holders (dc and microwave) are equipped with an electromagnetic coil for the application of a variable out-of-plane magnetic field during the measurements.
\end{itemize}

\subsection{Experimental setups}
\label{app:setups}
\begin{figure*}
	\includegraphics{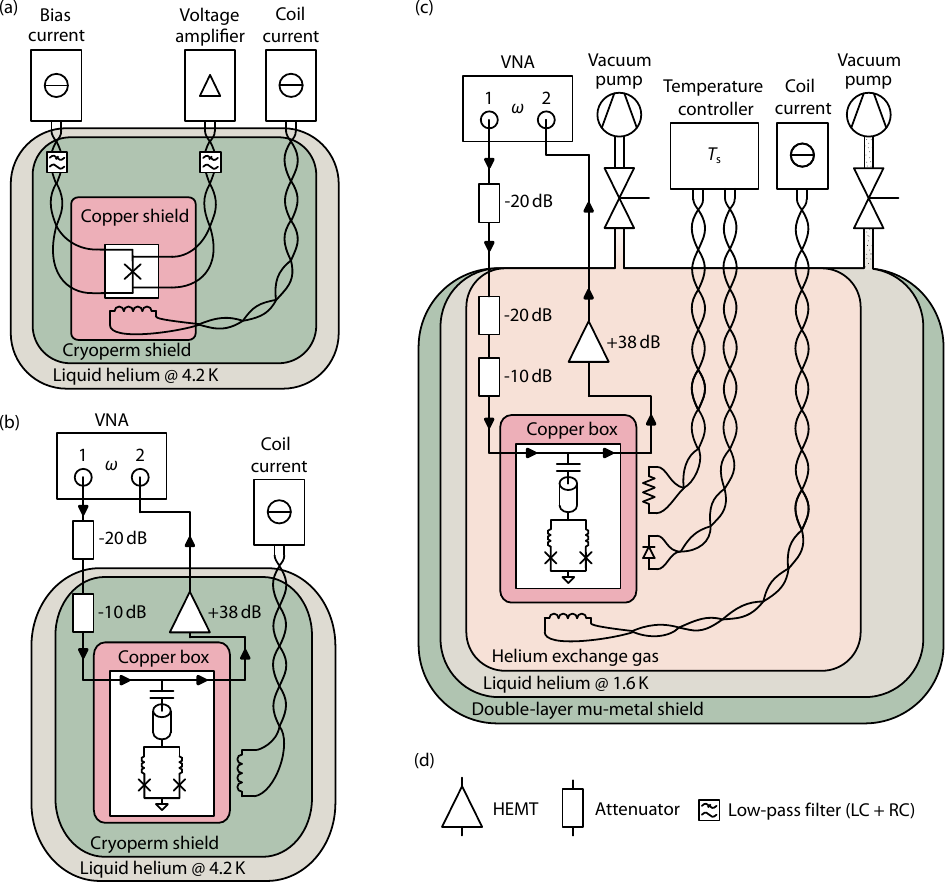}
	\titlecaption{Experimental setup schematics.}{%
		\sublabel{a}~Schematic of the experimental setup used for dc characterization of single bJJs in liquid helium~(data in Fig.~\ref{fig:figure2}).
		\sublabel{b}~Schematic of the experimental setup used for microwave characterization of SQUID cavities in liquid helium (data in Figs.~\ref{fig:figure3} and~\ref{fig:figure4}) and for pre-characterization before bJJ irradiation (data in Fig.~\ref{fig:figure1}); for the latter without HEMT, input line attenuation, or magnetic field.
		\sublabel{c}~Schematic of the experimental setup used for temperature-controlled microwave experiments in vacuum, more specifically in low-pressure helium exchange gas (data in Fig.~\ref{fig:figure5}).
		\sublabel{d}~Symbol legend for \sublabel{a}\textendash\sublabel{c}.
		Details regarding all setups are given in the text, cf.~Appendix\ref{app:setups}.
	}
	\vspace{+2pt}
	\label{fig:figure6}
\end{figure*}

\vspace{-2mm}
In total, three similar but separate experimental setups have been used for this study, all of which are schematically shown in Fig.~\ref{fig:figure6}.

For the dc IVC experiments of the individual bJJs and the characterization of their ion-dose dependence discussed in Fig.~\ref{fig:figure2}, a dedicated dc dipstick for measurements in liquid helium transport dewars was used.
The sample was directly immersed in liquid helium for the transport experiments, and therefore the measurement temperature was $T_\mathrm{s} = \SI{4.2}{\kelvin}$.
The dipstick is equipped with twelve pairs of twisted copper wires for IVC measurements, all of which are cryogenically low-pass filtered near the device by a combination of RC and LC filters.
For magnetic and high-frequency shielding as well as mechanical stability, the sample space is surrounded by two long, concentric cylindrical tubes (inner diameter $\qty{\sim 40}{\milli\meter}$), the inner of which is made of copper and the outer of which is made of Cryoperm.
Both have a closed bottom with a small hole to allow liquid helium to flow in and out.
Directly attached to the sample holder is a small electromagnetic coil for the application of an out-of-plane magnetic field, which is connected to a separate pair of twisted copper wires.
Currents for both sample and coil are provided by homemade battery-powered low-noise current sources and voltages are detected using a homemade low-noise amplifier and a computer-controlled analog-to-digital converter.
Data acquisition and current control was done using the software package \textit{GoldExi}.
Most of the microwave experiments (Figs.~\ref{fig:figure1}, \ref{fig:figure3} and \ref{fig:figure4}) have been conducted in a similar setting inside a liquid helium transport dewar, but with a dipstick designed for low-noise microwave experiments.
In this case as well, the sample was directly immersed in liquid helium and the measurement temperature was $T_\mathrm{s} = \SI{4.2}{\kelvin}$.
The dipstick is equipped with four coaxial lines and a few twisted-pair copper wires.
One of the twisted-pairs was connected to the electromagnetic coil attached to the sample box.
Two of the coaxial lines were used as input and output for the measurement of $S_{21}$ with a VNA.
For the measurements with bJJs in the resonators, the input line is attenuated by $\qty{30}{\decibel}$ plus cable attenuation, and the output line is equipped with a cryogenic high-electron-mobility-transistor (HEMT) amplifier for maximized signal-to-noise ratio.
For the pre-characterization without bJJs a nearly identical dipstick without attenuators, HEMT and magnetic coil was used, which is not shown here for the sake of simplicity.
The chip itself is packaged into a radiation-tight copper housing and the complete sample space is surrounded by a $\qty{\sim 30}{\centi\meter}$ long cylindrical tube (inner diameter $\qty{\sim 48}{\milli\meter}$) of Cryoperm for magnetic shielding.
The dc current for the coil is generated by a low-noise and battery-powered current source (identical to the ones used for the dc experiments), and $S_{21}$ traces are taken using a VNA.
Current source and VNA are controlled via \textit{Python}-based measurement scripts.
The final experiment regarding the temperature dependence of the Josephson cavities was conducted in the vacuum compartment of a more sophisticated microwave cryostat.
Through a combination of pumping on the cryostat liquid-helium bath, some helium exchange gas inside the vacuum compartment with the sample, and a feedback control loop using a temperature sensor and a heating resistor, both mounted close to the sample, this setup enables temperature control between $T_\mathrm{s}^\mathrm{min} \sim \qty{2}{\kelvin}$ (without the HEMT amplifier even down to $\qty{\sim 1.6}{\kelvin}$) and $T_\mathrm{s}^\mathrm{max} \sim \qty{60}{\kelvin}$ with a temperature stability $\Delta T_\mathrm{s} < \qty{1}{\milli\kelvin}$.
Here, we limited our experiments to $T_\mathrm{s} \in\left[2.1, 14\right]\,$K, since for higher $T_\mathrm{s}$ the internal linewidth of the resonator got too large to conduct a useful data analysis.
The input/output lines in the cryostat insert are similar to the ones in the transport dewar experiment (four coaxial lines, one input with $\qty{30}{\decibel}$ attenuation, one output with a cryogenic HEMT, multiple twisted-pair copper wires).
However, instead of a small Cryoperm magnetic shield like in the other setups, the complete cryostat is surrounded by a double-layer mu-metal shield at room-temperature for magnetic shielding.
VNA, coil current and temperature are all controlled via \textit{Python}-based measurement scripts from a single PC; vacuum pumps and HEMT are operated manually.
\vspace{-3mm}
\subsection{Background correction and resonance fitting}
\label{app:fitting}
\vspace{-2mm}
In general, our data analysis precedure for the $S_{21}$ scattering matrix elements can be divided into two steps, a background-correction with an experimentally obtained background transmission and a \textit{Python}-based fitting routine of the background-corrected data, from which we obtain $\omega_0$, $\kappa_0$ and $\kappa_\mathrm{ext}$~(or $\omega_\mathrm{c}, \kappa_\mathrm{c}$, and $\kappa_\mathrm{ext, c}$ before bJJ introduction).
The output of the fitting routine is then used for a second, fit-based background-correction, but this second correction is only for cross-checking the quality of the fit and for plotting purposes.
The first step --- the data-based background correction --- is mainly useful for large-linewidth resonances and therefore in this work we are not using it for the resonance datasets measured before writing the bJJs (cf.~Fig.~\ref{fig:figure1}), but we do for all datasets with bJJs in the cavities.
The basic idea is that we measure $S_{21}$ also in absence of the cavity resonance, which can either be achieved by using a high power VNA probe tone, such that the cavity dip is suppressed due to the JJ nonlinearities, or by tuning the resonance far away using bias flux.
Here, we implemented the second method, meaning that we stitch together the overall background from at least two VNA traces, each time removing the part with the resonance and filling the gap with the data from another bias flux value, where the resonance has moved somewhere else.
With this approach we get a complex-valued background trace $S_{21}^\mathrm{bg}$ for each individual experiment (frequency range, number of points etc.~need to match the actual data).
Then, we compute $S_{21}^\mathrm{bgc} = S_{21}^\mathrm{rd}/S_{21}^\mathrm{bg}$, where the superscripts \enquote*{rd} and \enquote*{bgc} stand for \enquote*{raw data} for \enquote*{background-corrected}, respectively.
At this stage most of the resonances already look very symmetric and the remaining background is close to $1$.
Nevertheless we want to be as careful as possible and therefore fit the background-corrected dataset $S_{21}^\mathrm{bgc}$ with the very general function
\begin{equation}\label{eqn:S21_fit}
	S_{21}^\mathrm{fit} = \left(a_0 + a_1\omega + a_2\omega^2\right)\left[1 - \frac{\kappa_\mathrm{ext}\e^{\i\theta}}{\kappa_0 + 2\i\left(\omega - \omega_0\right)}\right]\e^{\i\left(\phi_0 + \phi_1\omega\right)}.
\end{equation}
The first factor (second-order polynomial in $\omega$) considers a possible variation of the background transmission amplitude with angular frequency $\omega$ and with $a_0, a_1$, and $a_2$ as fit parameters.
The last factor considers a similar (but only first-order) frequency-dependence of the background phase with $\phi_0$ and $\phi_1$ as fit parameters.
The central factor represents almost the ideal cavity response, but contains an additional Fano-rotation angle $\theta$, which can be interpreted as originating from interferences between different signal paths within the setup.
This Fano-angle $\theta$, the external decay rate $\kappa_\mathrm{ext}$, the internal decay rate $\kappa_\mathrm{int} = \kappa_0 - \kappa_\mathrm{ext}$ and the cavity resonance frequency $\omega_0$ are the remaining fit parameters.
Once they are all known, the ideal cavity response function
\begin{equation}
	S_{21}^\mathrm{ideal} = 1 - \frac{\kappa_\mathrm{ext}}{\kappa_0 + 2\i\left(\omega - \omega_0\right)}
\end{equation}
can be reconstructed by dividing off the first and third factors and rotating the resonance circle around its anchor point by~$-\theta$.
Note that the fit is performed in three steps due to the large amount of fit parameters.
In a first step, we cut out the resonance dip and fit only the background.
Then, we divide the data by this first background fit and fit the remaining resonance with the central factor of Eq.~(\ref{eqn:S21_fit}) only.
As a last step, we fit the original data again with the complete expression and use as starting values for all fit parameters the ones we obtained from the first two partial fits.

\vspace{-3mm}
\subsection{From CPW cavity to LC circuit without JJs}
\label{app:LC_noJJ}
\vspace{-2mm}
\begin{table*}
	\titlecaption{Designed, simulated and experimental parameters of the three cavities.}
	{The cavity length $l$ was chosen by design, the equivalent capacitance is calculated via $C_\mathrm{p} = C' l /2$ with the design value for $C' = \qty{147}{\pico\farad\per\meter}$. The parameters $\omega_\mathrm{c}$, $\kappa_\mathrm{ext, c}$ and $\kappa_\mathrm{int, c}$ were determined experimentally at $T_\mathrm{s} = \qty{4.2}{\kelvin}$. The coupling capacitance $C_\mathrm{c}$ is calculated from $\kappa_\mathrm{ext, c}$, $\omega_\mathrm{c}$, $C_\mathrm{p}$, and the feedline impedance $Z_0 = \qty{50}{\ohm}$. The inductance $L_\mathrm{p}$ is obtained from $\omega_\mathrm{c}$ and $C_\mathrm{tot} = C_\mathrm{p} + C_\mathrm{c}$. The penetration depth $\lambda_\mathrm{L}$ and the theoretical value for the loop inductance $L_\mathrm{loop}$ we found from numerical simulations with the software packages \textit{Sonnet} and \textit{3D-MLSI}~\cite{Khapaev2001}, respectively. Finally, the values for $L_\mathrm{r} = L' l/2$ are calculated from $C_\mathrm{tot}$ and $\omega_\mathrm{c}$.}
	\begin{tabular}{ c  c  c  c  c  c  c  c  c  c  c }
		\toprule
		Cavity & $l\,$($\qty{}{\milli\meter}$) & $C_\mathrm{p}\,$($\qty{}{\femto\farad}$) & $\omega_\mathrm{c}/2\pi\,$($\qty{}{\giga\hertz}$) & $\kappa_\mathrm{ext, c}/2\pi\,$($\qty{}{\mega\hertz}$) & $\kappa_\mathrm{int, c}/2\pi\,$($\qty{}{\mega\hertz}$) & $C_\mathrm{c}\,$($\qty{}{\femto\farad}$) & $L_\mathrm{p}\,$($\qty{}{\nano\henry}$) & $\lambda_\mathrm{L}\,$($\qty{}{\nano\meter}$) & $L_\mathrm{loop}\,$($\qty{}{\pico\henry}$) & $L_\mathrm{r}\,$($\qty{}{\nano\henry}$) \\
		\midrule
		R1 & 5.772 & 423 & 4.869 & 2.731 & 1.454 & 18 & 2.42 & 230 & 36 & 1.50 \\
		R2 & 4.812 & 353 & 5.810 & 6.145 & 2.394 & 20 & 2.01 & 230 & 36 & 1.24 \\
		R3 & 4.141 & 303 & 6.727 & 7.122 & 2.633 & 17 & 1.74 & 230 & 36 & 1.08 \\
		\bottomrule
	\end{tabular}
	\label{tab:table1}
\end{table*}
Before we introduce the He-bJJs, we characterize the cavities as discussed in Fig.~\ref{fig:figure1}.
To model the junctionless standing-wave cavities as simple LC oscillators, which gives some intuition about their properties and allows us to derive useful quantities, we consider the input impedance of a lossless short-terminated transmission line (including small losses is not complicated but does not add much insight here)
\begin{equation}
	Z_\mathrm{in} = \i Z_1 \tan{\beta l}
\end{equation}
where $Z_1 = \sqrt{L'/C'}$ is the CPW characteristic impedance, $l$ is the cavity length, and $\beta = \omega/v_\phi$ is the phase constant with the phase velocity $v_\phi = 1/\sqrt{L'C'}$; $L'$ and $C'$ represent the inductance and capacitance per unit length of the cavity CPW, respectively.
The quarter-wave resonance we are interested in is the one around $\beta_\mathrm{b} l = \pi/2$, where $Z_\mathrm{in}$ diverges, and where $\beta_\mathrm{b} = \omega_\mathrm{b}/v_\phi$ is defined by the bare cavity resonance frequency $\omega_\mathrm{b}$ before junction writing and without the coupling capacitance.
We find the first-order Taylor series of the input admittance at that frequency as
\begin{align}
	\frac{1}{Z_\mathrm{in}} & \approx \frac{\i}{Z_1}\left(\beta l - \beta_\mathrm{b} l\right) \\
	& = \i\frac{l}{Z_1 v_\phi}\left(\omega - \omega_\mathrm{b}\right) \\
	& = \i C' l \left(\omega - \omega_\mathrm{b}\right)
\end{align}
which is completely equivalent to the first-order Taylor expansion of the input admittance of a parallel LC circuit around its resonance frequency, if we set $C_\mathrm{p} = C' l/2$ with the lumped element capacitance of the LC circuit $C_\mathrm{p}$.
From the resonance frequency $\omega_\mathrm{b} = 1/\sqrt{L_\mathrm{p}C_\mathrm{p}}$, we can then also derive the equivalent lumped element inductance $L_\mathrm{p} = \frac{8}{\pi^2}L' l$.
We summarize the lumped element equivalents we got from this analogy as
\begin{equation}
	C_\mathrm{p} = \frac{C' l}{2}, ~~~~~ L_\mathrm{p} = \frac{8}{\pi^2}L' l, ~~~~~ \omega_\mathrm{b} = \frac{1}{\sqrt{L_\mathrm{p}C_\mathrm{p}}}.
\end{equation}
As a next step, we include the coupling capacitance $C_\mathrm{c}$ and get for the input impedance
\begin{equation}
	Z_\mathrm{in} = \frac{1}{\i\omega C_\mathrm{c}} + \i Z_1  \tan{\beta l}.
\end{equation}
The coupling capacitance has transformed the effective parallel LC resonance into a series resonance, which leads to the resonance condition $Z_\mathrm{in} = 0$, and we are interested in its behaviour around $\beta l \approx \pi/2$.
Replacing the $\tan$-function by its first-order Taylor series (only a good approximation for high-$Q_\mathrm{ext}$ cavities) leads to
\begin{align}
	Z_\mathrm{in} & \approx -\i\left[ \frac{1}{\omega C_\mathrm{c}} + \frac{Z_1}{\frac{l}{v_\phi}\left(\omega - \omega_\mathrm{b}\right)}\right] \\
	& = -\i\left[ \frac{1}{\omega C_\mathrm{c}} + \frac{1}{2C_\mathrm{p}\left(\omega - \omega_\mathrm{b}\right)}\right]
\end{align}
which vanishes for the new, coupled resonance frequency
\begin{equation}
	\omega_\mathrm{c} = \frac{\omega_\mathrm{b}}{1 + \frac{C_\mathrm{c}}{2 C_\mathrm{p}}}.
\end{equation}
For $C_\mathrm{c} \ll C_\mathrm{p}$, this is again approximately identical to the LC circuit case
\begin{equation}
	\omega_\mathrm{c} = \frac{1}{\sqrt{L_\mathrm{p}C_\mathrm{tot}}}, ~~~~~ C_\mathrm{tot} = C_\mathrm{p} + C_\mathrm{c},
\end{equation}
i.e., to a capacitively side-coupled parallel LC circuit.
Hence, we can also use the expression for the external decay rate of the capacitively side-coupled parallel LC circuit which is
\begin{align}
	\kappa_\mathrm{ext} & = \frac{Z_0 C_\mathrm{c}^2}{2L_\mathrm{p}\left(C_\mathrm{p} + C_\mathrm{c}\right)^2} \\
	& \approx \frac{\omega_\mathrm{c}^2C_\mathrm{c}^2 Z_0}{2C_\mathrm{p}}.
	\label{eqn:kext}
\end{align}
The latter expression enables us to estimate $C_\mathrm{c}$ based on the knowledge of feedline impedance $Z_0$, capacitance $C_\mathrm{p}$ and the experimentally obtained quantities $\omega_\mathrm{c}$ and $\kappa_\mathrm{ext}$.
Note that in the form Eq.~(\ref{eqn:kext}) it is valid both before and after bJJ introduction (with the replacement $\omega_\mathrm{c} \rightarrow \omega_0$), and so we omit the subscript \enquote*{c} in $\kappa_\mathrm{ext}$ here.
\vspace{-3mm}
\subsection{Device parameters without JJs}
\label{app:dev_parameters}
\vspace{-2mm}
In this appendix section, we summarize all relevant pre-bJJ parameters for the three CPW cavities discussed throughout this work.
All three are based on identical CPW structures with center conductor width $S = \qty{19.5}{\micro\meter}$ and gap width $W = \qty{9.5}{\micro\meter}$, which corresponds to a capacitance per unit length $C' = \qty{147}{\pico\farad\per\meter}$ and inductance per unit length $L' \approx \qty{518}{\nano\henry\per\meter}$.
The latter is obtained from the experimentally determined $\omega_\mathrm{c}$ and $\kappa_\mathrm{ext, c}$ as well as from the calculated $C'$, $l$ and feedline impedance~$Z_0$.
Hence, we obtain a characteristic impedance $Z_1 \approx \qty{60}{\ohm}$ and a phase velocity $v_\phi \approx \qty{1.15e8}{\meter\per\second}$.
Note, that $L'$ contains both geometric and kinetic contributions $L_\mathrm{g}'$ and $L_\mathrm{k}'$, and from an analytical calculation of the geometric inductance we find $L_\mathrm{g}' = \qty{408}{\nano\henry\per\meter}$, i.e., $L_\mathrm{k}' \approx \qty{110}{\nano\henry\per\meter}$.
Table~\ref{tab:table1} lists first the design parameters for length $l$ and the resulting capacitance $C_\mathrm{p}$ for all three resonators, then the experimentally determined $\omega_\mathrm{c}$, $\kappa_\mathrm{ext, c}$ and $\kappa_\mathrm{int, c}$, and finally the values for $C_\mathrm{c}$ and $L_\mathrm{p}$ as determined from all the other quantities.
To complete the cavity parameters, we determine the London penetration depth $\lambda_\mathrm{L}$ by simulating R1 with the software package \textit{Sonnet} and varying the surface inductance $L_\mathrm{s}$ until measured and simulated resonance frequencies agree.
We find best agreement for $L_\mathrm{s} = \mu_0 \lambda_\mathrm{L}\coth\left(d/\lambda_\mathrm{L} \right) = \qty{1.35}{\pico\henry}/\square$, which corresponds to a London penetration depth $\lambda_\mathrm{L} \approx \qty{230}{\nano\meter}$ in good agreement with other reports on YBCO thin film circuits~\cite{Zaitsev2002, MartinezPerez2017, Keenan2021}.
Using this value for $\lambda_\mathrm{L}$, we numerically calculate the SQUID loop inductance $L_\mathrm{loop} \approx \qty{36}{\pico\henry}$ using the software package \textit{3D-MLSI}~\cite{Khapaev2001}.
As an additional check, we also calculate the inductance $L_\mathrm{r} = L' l/2$ using \textit{3D-MLSI} with $\lambda_\mathrm{L} = \qty{230}{\nano\meter}$, and we find good agreement with the experimental values, e.g., for R1 we obtain $\qty{1.54}{\nano\henry}$ as compared to the experimentally extracted value of $\qty{1.50}{\nano\henry}$.
All the parameters above and in Table~\ref{tab:table1} are given at $T_\mathrm{s} = \qty{4.2}{\kelvin}$.
In principle, we would have to consider that for higher temperatures, such as those discussed in Fig.~\ref{fig:figure5}, the penetration depth will increase, which would lead to temperature-dependent inductances $L_\mathrm{r}$ and $L_\mathrm{loop}$ as well as to a temperature-dependent resonance frequency $\omega_\mathrm{c}$.
It is not exactly known though, what the temperature-dependence for $\lambda_\mathrm{L}$ in our \mbox{M-type} YBCO is.
Previous experiments with YBCO films of similar quality have found phenomenological dependences of the form~\cite{Zaitsev2002, MartinezPerez2017, Keenan2021}
\begin{equation}
	\lambda_\mathrm{L}(T_\mathrm{s}) = \lambda_\mathrm{L}(0)\left[1 - \left(\frac{T_\mathrm{s}}{T_\mathrm{c}}\right)^{p_1}\right]^{-\frac{1}{p_2}}
\end{equation}
with typically $p_1\gtrsim 2$ and $p_2\gtrsim 2$, in particular for films with large $\lambda_\mathrm{L}(\qty{4.2}{\kelvin})$ as is the case here.
If we take exemplarily \mbox{$p_1 = 2$} and \mbox{$p_2 = 3$} as in Ref.~\cite{MartinezPerez2017} and \mbox{$T_\mathrm{c} = \qty{87}{\kelvin}$}~(value according to the datasheet from Ceraco), we get \mbox{$\lambda_\mathrm{L}(\qty{2.1}{\kelvin}) \approx \qty{229.8}{\nano\meter}$} and \mbox{$\lambda_\mathrm{L}(\qty{14}{\kelvin}) \approx \qty{231.8}{\nano\meter}$}.
This is equivalent to total inductance changes ${\lesssim }\qty{0.4}{\percent}$ for $L_\mathrm{r}$ (kinetic inductance contributes roughly $\qty{20}{\percent}$ to $L_\mathrm{r}$), and to resonance frequency changes~${\lesssim}\qty{0.2}{\percent}$.
For R1 this would translate to a temperature-induced frequency-shift $\delta\omega_\mathrm{c}(T_\mathrm{s}) \lesssim 2\pi\times \qty{9}{\mega\hertz}$.
With a similar reasoning, we find $\delta L_\mathrm{loop}(T_\mathrm{s}) < \qty{1}{\pico\henry}$.
The frequency-shift values could also be considerably smaller (e.g.~for $p_1 > 2$ as found in Ref.~\cite{Keenan2021}), and might be compensated for by other effects like He exchange gas condensing onto the chip for the lowest temperatures or a slightly temperature-dependent permittivity of the MgO substrate.
Therefore, and due to the unfortunate lack of a junction-free reference cavity, we assume for the sake of simplicity that $L_\mathrm{r}$, $L_\mathrm{loop}$ and $\omega_\mathrm{c}$ are not significantly dependent on $T_\mathrm{s}$ and treat them as constants.
Also, we do not explicitly consider differences in $C'$ and $\omega_\mathrm{c}$ for measurements in liquid He and measurements in low-density helium gas, which based on the relative permittivity of liquid helium $\epsilon_\mathrm{He} \approx 1.056$ are also on the order of $\qty{0.4}{\percent}$ for $C'$ and $\qty{0.2}{\percent}$ for $\omega_\mathrm{c}$.
For each experiment, however, we will let $\omega_\mathrm{c}$ be a temperature-independent fit parameter, see~Appendix\ref{app:Befaftcomp}.
For the experiments in liquid helium, we attribute the change of $\omega_\mathrm{c}$ to an increased $L_\mathrm{r}$, while $C'$ is assumed constant, and in helium gas to a decreased $C'$, while $L_\mathrm{r}$ is constant.
\vspace{-3mm}
\subsection{From CPW cavity to LC circuit with JJs}
\label{app:LC_wJJ}
\vspace{-2mm}
To find the resonance frequency, when the cavity is not shorted at its SQUID end but connected to ground via a flux-tunable Josephson inductance, we analyze the cavity from the SQUID side.
This means instead of resembling a parallel LC circuit around resonance as in~Appendix\ref{app:LC_noJJ}, the bare cavity around its $\lambda/4$-mode will be analogous to a series LC circuit.
We start again by omitting the coupling capacitance and follow the approach described in Refs.~\cite{Uhl2024a, Pogorzalek2017}, where in order to find the resonance frequency we set the input impedance of the cavity equal to that of the Josephson inductances with opposite sign
\begin{equation}
	\i Z_1 \cot{\beta l} = \i\omega L_\mathrm{S}.
\end{equation}
Here, $L_\mathrm{S} = L_\mathrm{J}/2$ is the Josephson part of the SQUID inductance for identical JJs and $L_\mathrm{J} = \Phi_0/2\pi I_0\cos\left(\pi\Phi/\Phi_0 \right)$ is the flux-dependent inductance of a single junction.
The contribution from the loop inductance is not explicitly included here, since it was already part of the cavity before introducing the bJJs.
To find the resonance frequency $\omega_\mathrm{s}$, we Taylor-approximate the $\cot$-function around $\beta l = \pi/2$ and get
\begin{equation}
	-\frac{Z_1 l }{v_\phi}\left(\omega_\mathrm{s} - \omega_\mathrm{b}\right) = \omega_\mathrm{s} L_\mathrm{S}
	\label{eqn:Taylor_cot}
\end{equation}
or
\begin{align}
	\omega_\mathrm{s} & = \frac{\omega_\mathrm{b}}{1 + \frac{L_\mathrm{S}v_\phi}{Z_1 l}} \\
	& = \frac{\omega_\mathrm{b}}{1 + \frac{L_\mathrm{S}}{L' l}}.
\end{align}
We see a certain symmetry when comparing this to the result of the coupling capacitance, and this time the lumped element equivalents of the bare cavity are
\begin{equation}
	L_\mathrm{r} = \frac{L' l}{2}, ~~~~~ C_\mathrm{r} = \frac{8}{\pi^2}C' l, ~~~~~ \omega_\mathrm{b} = \frac{1}{\sqrt{L_\mathrm{r}C_\mathrm{r}}}.
\end{equation}
The resonance frequency of the cavity with the Josephson termination is
\begin{equation}
	\omega_\mathrm{s} = \frac{\omega_\mathrm{b}}{1 + \frac{L_\mathrm{S}}{2 L_\mathrm{r}}}.
\end{equation}
As a final step, we consider the impact of the coupling capacitance at the far end with this approach.
The input impedance seen from the Josephson termination is
\begin{align}
	Z_\mathrm{in} & = -\i Z_1\frac{1 - Z_1\omega C_\mathrm{c}\tan{\beta l}}{Z_1\omega C_\mathrm{c} + \tan{\beta l}} \\
	& = -\i Z_1\frac{\cot{\beta l} - Z_1\omega C_\mathrm{c}}{1 + Z_1\omega C_\mathrm{c}\cot{\beta l}}
\end{align}
and we replace again the $\cot$-terms in this expression by their first-order Taylor series
\begin{align}
	Z_\mathrm{in} & \approx \i Z_1\frac{\frac{l}{v_\phi}\left(\omega - \omega_\mathrm{b}\right) + Z_1\omega C_\mathrm{c}}{1 - Z_1\omega C_\mathrm{c}\frac{l}{v_\phi}\left(\omega - \omega_\mathrm{b}\right)} \\
	& \approx \i Z_1\left[\frac{l}{v_\phi}\left(\omega - \omega_\mathrm{b}\right) + Z_1\omega C_\mathrm{c}\right].
\end{align}
Then, we equate this to $\i\omega L_\mathrm{S}$ to find the resonance frequency of the capacitively coupled Josephson cavity $\omega_0$ via
\begin{equation}
	Z_1\left[\frac{l}{v_\phi}\left(\omega_0 - \omega_\mathrm{b}\right) + Z_1\omega_0 C_\mathrm{c}\right] = -\omega_0 L_\mathrm{S}
\end{equation}
and end up with
\begin{eqnarray}
	\omega_0 & = & \frac{\omega_\mathrm{b}}{1 + \frac{C_\mathrm{c}}{C' l} + \frac{L_\mathrm{S}}{L' l}} \\
	 & = & \frac{\omega_\mathrm{b}}{1 + \frac{C_\mathrm{c}}{2 C_\mathrm{p}} + \frac{L_\mathrm{S}}{2 L_\mathrm{r}}} \\
	 & \approx & \frac{\omega_\mathrm{c}}{1 + \frac{L_\mathrm{S}}{2 L_\mathrm{r}}}.
	 \label{eqn:w0_app}
\end{eqnarray}
The values for $L_\mathrm{r} = \pi^2/16C_\mathrm{tot}\omega_\mathrm{c}^2$ are also added to Table~\ref{tab:table1}.
%

\subsection{Flux-tuning by SQUID}
\label{app:fluxSQUID}
%
To model the tuning of the cavities with bias flux $\Phi_\mathrm{b}$ through the SQUIDs, we first need the relation between the total flux in the SQUID $\Phi$ and the bias flux $\Phi_\mathrm{b}$, which for symmetric SQUIDs (JJs with equal critical currents) and no transport current is given by
\begin{equation}
	\frac{\Phi}{\Phi_0} = \frac{\Phi_\mathrm{b}}{\Phi_0} - \frac{\beta_L}{2}\sin\left(\pi\frac{\Phi}{\Phi_0} \right)
	\label{eqn:flux}
\end{equation}
with the flux quantum $\Phi_0 \approx \qty{2.068e-15}{\tesla\meter\squared}$ and the screening parameter $\beta_L = L_\mathrm{loop}/\pi L_\mathrm{J0}$, where $L_\mathrm{J0} = \Phi_0/2\pi I_0$ is the single-junction sweetspot inductance and $I_0$ is the single-junction critical current.
On the other hand, the low-amplitude Josephson inductance of a single bJJ is given by
\begin{equation}
	L_\mathrm{J} = \frac{L_\mathrm{J0}}{\cos\left(\pi\frac{\Phi}{\Phi_0}\right)}
\end{equation}
and so an explicit expression for the flux-dependent cavity resonance frequency is
\begin{equation}
	\omega_0 = \frac{\omega_\mathrm{c}}{1 + \frac{L_\mathrm{J0}}{4L_\mathrm{r}\cos\left(\pi\frac{\Phi}{\Phi_0}\right)}}
\end{equation}
which we can then use -- while simultaneously solving Eq.~(\ref{eqn:flux}) numerically -- to fit the flux-tuning data points $\omega_0(\Phi_\mathrm{b})$ obtained from the experiment.
We do this with $L_\mathrm{loop}$ and $C_\mathrm{tot}$ as fixed parameters obtained from the data analysis before junction writing, and with $I_0$ as a free fit parameter, which then determines $\tilde{\omega}_\mathrm{c}$ and $\tilde{L}_\mathrm{r} = \pi^2/16C_\mathrm{tot}\tilde{\omega}_\mathrm{c}^2$ (in the liquid helium measurements) or alternatively $\tilde{\omega}_\mathrm{c}^*$ and $\tilde{C}_\mathrm{tot} = \pi^2/16L_\mathrm{r}\tilde{\omega}_\mathrm{c}^{*2}$ (in the temperature-dependent measurements).
We use the tildes (and the asterisk) here to indicate possibly different values of $\tilde{\omega}_\mathrm{c}, \tilde{L}_\mathrm{r}$ or $\tilde{\omega}_\mathrm{c}^*, \tilde{C}_\mathrm{tot}$ obtained from the arc-fit and $\omega_\mathrm{c}, L_\mathrm{r}, C_\mathrm{tot}$ as calculated/measured/determined before the bJJ writing, see also Appendix\ref{app:Befaftcomp}.
For an asymmetric SQUID with two unequal bJJ critical currents $I_1$ and $I_2$, the flux equation has a slightly more complicated form~\cite{Paradkar2025}
\begin{equation}
	\frac{\Phi}{\Phi_0} = \frac{\Phi_\mathrm{b}}{\Phi_0} - \frac{\beta_L}{2}\frac{\left(1 - \alpha_I^2 \right)}{\sqrt{1 + \alpha_I^2 \tan^2\left(\pi\frac{\Phi}{\Phi_0} \right)}}\sin\left(\pi\frac{\Phi}{\Phi_0} \right)
	\label{eqn:flux_asy}
\end{equation}
where
\begin{equation}
	\alpha_I = \frac{I_1 - I_2}{I_1 + I_2}
\end{equation}
is the critical current asymmetry parameter, related to the average single-JJ critical current $I_0$ via
\begin{align}
	I_1 & = I_0(1 + \alpha_I) \\
	I_2 & = I_0(1 - \alpha_I).
\end{align}
The total Josephson SQUID inductance we find via
\begin{align}
	L_\mathrm{S} & = \frac{L_\mathrm{J1}L_\mathrm{J2}}{L_\mathrm{J1} + L_\mathrm{J2}} \\
	& = \frac{L_\mathrm{J0}}{2\cos\left( \pi\frac{\Phi}{\Phi_0}\right)\sqrt{1 + \alpha_I^2\tan^2\left( \pi\frac{\Phi}{\Phi_0}\right)}}
\end{align}
and hence, we can completely analogously to the symmetric case use these relations to fit the resonance frequency.
Strictly speaking the latter can only be used for negligible linear series inductance for the microwave currents inside the SQUID.
However, since the small arms of the SQUID only carry a fraction of the loop inductance, which is anyways much smaller than $L_\mathrm{J0}$ we believe this is a justified approximation here.

\begin{table*}
	\titlecaption{Cavity, junction and SQUID parameters after bJJ writing at the flux sweetspot.}
	{Resonance frequency $\tilde{\omega}_\mathrm{c}$ and resonator inductance $\tilde{L}_\mathrm{r}$ are fit parameters for the junctionless resonance, cf.~Eq.~(\ref{eqn:fluxarc_prime}). The sweetspot external decay rate $\kappa_{\mathrm{ext}0}$ is obtained from resonance fits around the flux sweetspot after bJJ writing. Critical current $I_0$, sweetspot Josephson inductance $L_\mathrm{J0} = \Phi_0/2\pi I_0$ and SQUID screening parameter $\beta_L = L_\mathrm{loop}/\pi L_\mathrm{J0}$ are obtained from the flux-tuning fits of each cavity. The Josephson sweetspot decay rate induced by writing the bJJs is calculated via Eq.~(\ref{eqn:kappa_J0_def}).}
	\begin{tabular}{ c  c  c  c  c  c  c  c }
		\toprule
		Cavity & $\tilde{\omega}_\mathrm{c}/2\pi\,$($\unit{\giga\hertz}$) & $\tilde{L}_\mathrm{r}\,$($\unit{\nano\henry}$) & $\kappa_{\mathrm{ext}0}/2\pi\,$($\unit{\mega\hertz}$) & $I_0\,$($\unit{\micro\ampere}$) & $L_\mathrm{J0}\,$($\unit{\pico\henry}$) & $\beta_L\,$ & $\kappa_{\mathrm{J}0}/2\pi\,$($\unit{\mega\hertz}$) \\
		\midrule
		R1 & 4.797 & 1.54 & 3.21 & 2.76 & 119 & 0.096 & 7.30 \\
		R2 & 5.699 & 1.29 & 4.81 & 1.90 & 173 & 0.066 & 20.5 \\
		R3 & 6.586 & 1.12 & 6.08 & 1.51 & 219 & 0.052 & 29.0 \\
		\bottomrule
	\end{tabular}
	\label{tab:table2}
\end{table*}
\vspace{2mm}
\subsection{Basic parameters before vs after bJJ writing}
\label{app:Befaftcomp}
%
When fitting the flux-tuning arcs as described above and in the main manuscript text, we obtain as one of the fit parameters the cavity resonance frequency $\tilde{\omega}_\mathrm{c}$ which is smaller than the actually measured frequency before bJJ writing $\omega_\mathrm{c}$ for all three cavities.
Similarly, we get a sweetspot external decay rate $\kappa_{\mathrm{ext}0}$ after junction writing that differs from the external decay rate before junction writing.
As an overview, we collect these two fit parameters for all three cavities in Table~\ref{tab:table2}.
Since we have determined the corresponding quantities $\tilde{\omega}_\mathrm{c}^*$ and $\kappa_{\mathrm{ext}0}^*$ only for R1 and they are only relevant for Fig.~\ref{fig:figure5}, we do not include them in the table or discussion explicitly.
Additionally, the table contains for each cavity the sweetspot critical junction current, the sweetspot single-bJJ Josephson inductance, the SQUID screening parameter, and the sweetspot junction decay rate
\begin{equation}
\kappa_{\mathrm{J}0} = \kappa_{\mathrm{int}0} - \kappa_\mathrm{int, c}\frac{\omega_{00}}{\omega_\mathrm{c}},
\label{eqn:kappa_J0_def}
\end{equation}
where $\kappa_\mathrm{int, c}$ is the internal decay rate from before bJJ writing and $\kappa_{\mathrm{int}0}$ its counterpart with bJJs.
The internal linewidth before bJJ writing is scaled by the resonance frequency ratio $\omega_{00}/\omega_\mathrm{c}$ here for consistency with the definition of $1/Q_{\mathrm{J}0}$, cf.~Eq.~(\ref{eqn:lossJ0_def}).
The impact of the scaling factor is negligibly small, though.
The external decay rates after bJJ writing deviate up to ${\sim}28\%$ from the ones before bJJ writing with one being larger~(R1), and the other two being smaller than before~(R2, R3).
This is an acceptable deviation and not unusual after de-mounting and re-mounting a sample to PCB and sample holder, since several contributions can change the apparent external decay rate.
First, the resonance frequency is lower after bJJ introduction, which is expected to decrease $\kappa_\mathrm{ext}$ even in a perfect setup, cf.~Eq.~(\ref{eqn:kext}), in our case ${\sim}\qty{7}{\percent}$ for R1, ${\sim}\qty{10}{\percent}$ for R2, and ${\sim}\qty{13}{\percent}$ for R3.
Additionally, there are two parasitic effects due to imperfections in the feedline and the microwave cabling: Partial standing waves on the feedline due to parasitic reflections and parallel transmission paths, bypassing the cavity, leading to Fano interferences~\cite{Rieger2023}.
The former causes the effective coupling capacitance to change as a function of the cavity position in the standing wave, and the latter introduces offsets to the complex transmission signal, which modifies the output of the fitting routine and falsely attributes parts of $\kappa_\mathrm{int}$ or $\kappa_\mathrm{J}$ to $\kappa_\mathrm{ext}$~\cite{Rieger2023}.
Both of these parasitic effects can furthermore be frequency-dependent.
The shift of $\tilde{\omega}_\mathrm{c}$ with respect to $\omega_\mathrm{c}$, which is around $\qty{2}{\percent}$ in our experiment, seems more systematic (all $\tilde{\omega}_\mathrm{c} < \omega_\mathrm{c}$) and likely has a different origin, although de-mounting and re-mounting can also shift the frequency by e.g.~modified interactions with box and chip modes.
One important further aspect is that five months had passed between the pre-characterization of the junction-free chip and the bJJ writing in the HIM, and ten more weeks between bJJ writing and experiments.
This is sufficient time for the YBCO film to slightly degrade throughinteraction with ambient water (despite the chip being stored in a nitrogen atmosphere) and/or oxygen out-diffusion~\cite{Mueller1991, Behner1993, Russek1994, Perez1995}, which might increase the kinetic inductance contributions and hence lower the resonance frequencies.
In repeated measurements on similar chips as the one presented here, just without HIM bJJs, we observed frequency downshifts over long time periods on the order of several $\qty{10}{\mega\hertz}$, which is on a similar order of magnitude but still somewhat smaller than the shifts we find here.
Further potential contributions could stem from slightly asymmetric SQUIDs (different critical currents of the two JJs in each SQUID) or from weakly non-sinusoidal current-phase-relations of the bJJs.
Since in particular the latter two would lead to multiple additional fit parameters without deepening the insights, we treat all of the shift being caused by a change of linear cavity inductance $L_\mathrm{r} \rightarrow \tilde{L}_\mathrm{r}$, and we will investigate the origin of the effect in dedicated devices in more detail in the future.
%

\subsection{Flux-tuning by single-junction field dependence}
\label{app:fluxFraun}
%
For external magnetic fields that correspond to more than a few $\Phi_0$ in the SQUID, we need to take the field-tuning of the individual junctions into account.
To do that exemplarily here for an ideal SIS junction, we follow the standard theoretical description of a short JJ in an external magnetic field parallel to the barrier of the junction.
The field is applied in $z$-direction and the long side of the junction extends in $y$-direction.
Assuming a negligible screening current density in the superconducting electrodes sufficiently far away from the bJJ barrier, we can then derive an expression for the gauge-invariant phase-difference $\delta$ at two neighboring points along the junction
\begin{equation}
	\delta(y + dy) - \delta(y) = \frac{2\pi}{\Phi_0}\oint\limits_\mathcal{C} \vec{A}\cdot d\vec{l}
\end{equation}
with the vector potential $\vec{A}$ and the path element $d\vec{l}$ along a closed path $\mathcal{C}$ from one electrode into the second and back.
Using Stokes' theorem we get
\begin{equation}
	\oint\limits_\mathcal{C} \vec{A}\cdot d\vec{l} = \int\limits_\mathcal{S} \vec{B}\cdot d\vec{s}
\end{equation}
where the area element $d\vec{s}$ runs over the area $\mathcal{S}$ enclosed by the path $\mathcal{C}$.
Assuming a homogeneous magnetic field $\vec{B} = \vec{B}_\mathrm{ext}$ inside the junction, the integral can be directly solved as
\begin{equation}
	\int\limits_\mathcal{S} \vec{B}_\mathrm{ext}\cdot d\vec{s} = B_\mathrm{ext} d_\mathrm{eff} dy
\end{equation}
where the effective thickness of the bJJ with actual barrier thickness $d_\mathrm{J}$ is given by
\begin{equation}
	d_\mathrm{eff} \approx d_\mathrm{J} + 2\lambda_\mathrm{L}.
\end{equation}
As a consequence, we obtain the field-induced and stiff phase gradient along the bJJ as
\begin{equation}
	\frac{d\delta}{dy} = \frac{2\pi}{\Phi_0}B_\mathrm{ext} d_\mathrm{eff}.
\end{equation}
With the critical current density $j_0$, the local Josephson relation for the current flow across the junction
\begin{equation}
	j_x = j_0\sin{\delta_y}
\end{equation}
with a position-dependent $\delta_y = \delta(y)$ can now be expressed as a function of magnetic field
\begin{equation}
	j_x = j_0\sin\left(\delta_0 + \frac{2\pi}{\Phi_0}B_\mathrm{ext} d_\mathrm{eff} y\right)
\end{equation}
where $\delta_0$ came into the equation as an integration constant and constitutes the dynamical variable in the presence of a net supercurrent through the junction.
The total supercurrent is now obtained via
\begin{equation}
	I_x = j_0 t_\mathrm{Y} \int\limits_{-a/2}^{a/2}\sin\left(\delta_0 + \frac{2\pi}{\Phi_0}B_\mathrm{ext} d_\mathrm{eff} y\right) dy
\end{equation}
where $a$ is the width of the junction in $y$-direction and $t_\mathrm{Y}$ is the thickness of the YBCO film (junction dimension in $z$-direction).
Integration leads to
\begin{eqnarray}
	I_x & = & I_0 \sin{\delta_0}\frac{\sin\left(\pi\frac{\Phi_\mathrm{J}}{\Phi_0} \right)}{\pi\frac{\Phi_\mathrm{J}}{\Phi_0}} \\
	& = & I_B \sin{\delta_0},
	\label{eqn:CPR_B}
\end{eqnarray}
i.e., to a standard JJ with a magnetic-field-dependent critical current
\begin{equation}
	I_B = I_0 \frac{\sin\left(\pi\frac{\Phi_\mathrm{J}}{\Phi_0} \right)}{\pi\frac{\Phi_\mathrm{J}}{\Phi_0}}
\end{equation}
and where
\begin{equation}
	\Phi_\mathrm{J} = B_\mathrm{ext} d_\mathrm{eff} a
\end{equation}
is the Josephson flux.
The minimum single-junction inductance of each flux arc then lies on the curve
\begin{equation}
	L_{\mathrm{J}B} = \frac{\Phi_0}{2\pi I_B}
\end{equation}
and the expression for the resonance frequency in case of a symmetric SQUID is
\begin{eqnarray}
	\omega_0 & = & \frac{\omega_\mathrm{c}}{1 + \frac{L_{\mathrm{J}B}}{4L_\mathrm{r}\cos\left(\pi\frac{\Phi}{\Phi_0}\right)}} \label{eqn:w0B} \\
	& = & \frac{\omega_\mathrm{c}}{1 + \frac{ L_\mathrm{J0}}{4L_\mathrm{r}\cos\left(\pi\varphi\right)}\frac{\pi\varphi_\mathrm{J}}{\sin{(\pi\varphi_\mathrm{J})}}}
\end{eqnarray}
with $\varphi_\mathrm{J} = \Phi_\mathrm{J}/\Phi_0$ and $\varphi = \Phi/\Phi_0$.
It can be easily generalized to the asymmetric SQUID if required, but since we only use the asymmetry in context of Fig.~\ref{fig:figure5} and with very few $\Phi_0$ in the SQUID, we do not do so explicitly here.
Note that Eqs.~(\ref{eqn:CPR_B}) and (\ref{eqn:w0B}) are quite general, as long as the junctions in the SQUID have a sinusoidal current-phase-relation.
Then, even inhomogeneities in the critical-current density or the local magnetic field will not alter the overall shape, but only the shape of $I_B$ and $L_{\mathrm{J}B}$.
Note furthermore that the above considerations are simplified for illustration, and the actual HIM bJJs used here do not exactly fulfill the approximations of this simple theory, in particular not a vanishing screening current-density in the leads.
The junctions are rather close to or in the nonlocal limit, and it has been shown that in this case the effective thickness, which defines the field of the first $I_B$ minimum, is given by~\cite{Clem2010}
\begin{equation}
	d_\mathrm{eff} \approx \frac{a^2}{1.5}.
\end{equation}
Other modifications in the nonlocal limit concern e.g.~the relative height of the side lobes and the periodicity of the minima, but since many other factors also generate anomalies in the interference pattern of YBCO He-bJJs, we will leave our considerations at the exemplary treatment of an ideal JJ.
\vspace{-1mm}
\subsection{Circuit-element-based loss factor}
\label{app:losses}
%
If we model each bJJ as a parallel combination of an inductance $L_\mathrm{J}$ and a resistor $R_\mathrm{J}$ as suggested by the RSJ model, we can calculate an expected loss factor as a function of these two parameters and the remaining cavity quantities.
Since we only discuss sweetspot values throughout the manuscript, it is sufficient to consider $L_{\mathrm{J}0}$; the resistance $R_\mathrm{J}$ we assume to be independent of field or flux.
In general $R_\mathrm{J} \neq R_\mathrm{n}$ with the voltage-state dc resistance $R_\mathrm{n}$, but we will assume them to be similar, which based on typical values for $R_\mathrm{n} \sim \qty{100}{\ohm}$ enables us to use the approximation $R_\mathrm{J}^2 \gg \omega_{00}^2L_{\mathrm{J}0}^2$.
To easily see its validity, the latter condition can be written as
\begin{equation}
	I_0^2 R_\mathrm{J}^2 \gg \frac{\omega_{00}^2 \Phi_{0}^2}{4\pi^2}
\end{equation}
using $L_\mathrm{J0} = \Phi_0/2\pi I_0$.
The quantity on the left hand side is ${\sim}V_\mathrm{c}^2$ with $V_\mathrm{c}$ the characteristic junction voltage, which for the current densities of our microwave devices is on the order of $\qty{250}{\micro\volt}$, cf.~Fig.~\ref{fig:figure7}.
This is equivalent to $I_0^2 R_\mathrm{J}^2 \sim V_\mathrm{c}^2 \approx \qty{6e-8}{\volt\squared}$.
The right hand side is between $\omega_{00}^2\Phi_0^2/4\pi^2 \approx \qty{1e-10}{\volt\squared}$ for R1 and $\qty{2e-10}{\volt\squared}$ for R3.
Neglecting the small inductance of each SQUID loop arm and assuming two different bJJs, we can write the input admittance of the two parallel SQUID arms around the sweetspot resonance frequency as
\begin{equation}
	\frac{1}{Z_\mathrm{RJ}} = \frac{1}{R_{\mathrm{J}1}} + \frac{1}{R_{\mathrm{J}2}} + \frac{1}{\i\omega_{00}L_{\mathrm{J}1}} + \frac{1}{\i\omega_{00}L_{\mathrm{J}2}}
\end{equation}
or
\begin{equation}
	Z_\mathrm{RJ} = \frac{\i\omega_{00}L_\parallel R_\parallel}{R_\parallel + \i\omega_{00}L_\parallel}
\end{equation}
with
\begin{equation}
	R_\parallel = \frac{R_{\mathrm{J}1}R_{\mathrm{J}2}}{R_{\mathrm{J}1} + R_{\mathrm{J}2}}, ~~~~~ L_\parallel = \frac{L_{\mathrm{J}1}L_{\mathrm{J}2}}{L_{\mathrm{J}1} + L_{\mathrm{J}2}}.
\end{equation}
Next, we transform this into a series equivalent of a resistor $R_\mathrm{J}^*$ and $L_\mathrm{J}^*$ by demanding
\begin{equation}
	\frac{\i\omega_{00}L_\parallel R_\parallel}{R_\parallel + \i\omega_{00}L_\parallel} = R_\mathrm{J}^* + \i\omega_{00}L_\mathrm{J}^*
\end{equation}
which leads to
\begin{equation}
	R_\mathrm{J}^* = \frac{\omega_{00}^2 L_\parallel^2 R_\parallel}{R_\parallel^2 + \omega_{00}^2 L_\parallel^2}, ~~~~~ L_\mathrm{J}^* = \frac{L_\parallel R_\parallel^2}{R_\parallel^2 + \omega_{00}^2 L_\parallel^2}.
\end{equation}
Our earlier found relation $R_\mathrm{J}^2 \gg \omega_{00}^2L_{\mathrm{J}0}^2$ is equivalent to $R_\parallel^2 \gg \omega_{00}^2 L_\parallel^2$ and hence we can approximate
\begin{equation}
	R_\mathrm{J}^* \approx \frac{\omega_{00}^2 L_\parallel^2}{R_\parallel}, ~~~~~ L_\mathrm{J}^* \approx L_\parallel.
\end{equation}
Finally, we can express the junction-related decay rate for the series equivalent circuit of the SQUID-shorted cavity as
\begin{equation}
	\kappa_{\mathrm{J}0} = \frac{R_\mathrm{J}^*}{L_\mathrm{r} + L_\mathrm{J}^*}
\end{equation}
or equivalently the loss factor as
\begin{align}
	\frac{1}{Q_{\mathrm{J}0}} & = \frac{R_\mathrm{J}^*}{\omega_{00}\left( L_\mathrm{r} + L_\mathrm{J}^*\right)}.
\end{align}
For two identical bJJs we get
\begin{equation}
	\frac{1}{Q_{\mathrm{J}0}} = \frac{\omega_{00}L_{\mathrm{J}0}^2}{R_\mathrm{J}\left( 2L_\mathrm{r} + L_{\mathrm{J}0}\right)}.
\end{equation}
\subsection{Additional data}
\label{app:add_data}
%
\subsubsection*{K.1:~Characteristic voltage of dc samples}
\begin{figure}
	\includegraphics{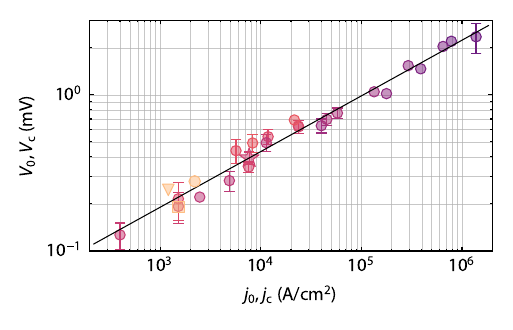}
	\vspace{-6mm}
	\titlecaption{Scaling of the characteristic voltage with critical current density.}{%
		Characteristic junction voltages $V_\mathrm{c} = I_\mathrm{c}R_\mathrm{n}$ of the dc bJJs and $V_0 = I_0R_\mathrm{J}$ of the cavity-embedded bJJs vs their critical current density $j_\mathrm{c} = I_\mathrm{c}/a_\mathrm{dc}$ and $j_0 = I_0/a$, respectively. Symbols are data extracted from experimental IVCs and microwave experiments, line is a fit ${\propto}j_\mathrm{c}^\alpha$ with fit parameter $\alpha = 0.36$. Color code of the dc results is identical to Fig.~\ref{fig:figure2}\sublabel{c}, star symbol corresponds to the bJJ discussed in Fig.~\ref{fig:figure2}(a, b), and the three yellow symbols correspond to the three cavities (R1:~circle, R2:~square, R3:~triangle). Absolute values range from $\qty{0.12}{\milli\volt}$ to $\qty{2.3}{\milli\volt}$. Error bars originate from the error in determining $R_\mathrm{n}$, for details see Appendix\ref{app:errors}; when the error is smaller than the symbol size, the error bars are omitted.
	}
	\vspace{-2mm}
	\label{fig:figure7}
\end{figure}

%
One important property of Josephson junctions is their characteristic voltage $V_\mathrm{c} = I_\mathrm{c} R_\mathrm{n}$, where $I_\mathrm{c}$ is the (measured) critical current and $R_\mathrm{n}$ is the voltage-state resistance of the IVCs in the approximately linear regime.
For cuprate Josephson junctions based on defect-induced junction barriers~(e.g.~through oxygen displacement), a universal scaling of the characteristic voltage ${\propto}j_\mathrm{c}^\alpha$ has been observed, where $\alpha \approx 0.5\pm0.1$ and $j_\mathrm{c}$ the critical current density of the bJJ~\cite{Hilgenkamp2002}.
Hence, observing such a scaling law in our junction ensemble would be a good indicator that all of our dc samples can be considered Josephson junctions and that the barriers are generated by helium-ion-induced lattice defects.
To determine the characteristic voltage we use the critical currents shown in Fig.~\ref{fig:figure2}\sublabel{c}, and additionally extract the (differential) voltage-state resistance $R_\mathrm{n}$ from the corresponding IVCs.
For a systematic and automatized extraction of $R_\mathrm{n} = \partial V / \partial I$, we fit small sections of the IVCs with a linear function around the current points given by $\pm2I_\mathrm{c}$, $\pm2.5I_\mathrm{c}$, and $\pm3I_\mathrm{c}$, except for the two IVCs with the highest $I_\mathrm{c}$, where we choose $\pm1.5I_\mathrm{c}$, $\pm1.75I_\mathrm{c}$, and $\pm2I_\mathrm{c}$; for larger currents the IVCs become rather nonlinear, both for small $I_\mathrm{c}$ and for large $I_\mathrm{c}$, likely due to heating effects and temperature-dependent~$R_\mathrm{n}$~\cite{Cybart2015}.
Finally, we calculate mean and standard deviation of the $R_\mathrm{n}$, and plot the characteristic voltage $ V_\mathrm{c} = I_\mathrm{c}R_\mathrm{n}$ in Fig.~\ref{fig:figure7}.
From the data in Fig.~\ref{fig:figure7} it is apparent that we indeed find a power-law scaling behaviour of $V_\mathrm{c}$ with $j_\mathrm{c}$.
We apply a fit to the data using the function~\cite{Mueller2019}
\begin{equation}
	V_\mathrm{c} = A_0 j_\mathrm{c}^\alpha
\end{equation}
with the free parameters $A_0$ and $\alpha$.
For the relevant power-law scaling parameter we find $\alpha = 0.36$, which is slightly smaller than, but still in reasonable agreement with the expectations.
Those results and the (compared to the critical current) small scattering of the $V_\mathrm{c}$ values around the fit curve indicate not only that we have defect-based Josephson junctions, but also that the scattering of critical currents for a fixed nominal dose is not related to variations in the underlying Josephson mechanism~(constriction-type micro-shorts e.g.~would likely scale differently~\cite{Schmid2025}), but rather to variations of their characteristics from one junction to the next, caused by e.g.~slowly fluctuating beam currents, a drifting beam focus size, or maybe variations in thin film properties.
Strikingly, also the values for $V_0 = R_\mathrm{J}I_0$ that we extracted from the three microwave devices using sweetspot inductances $L_{\mathrm{J}0} \propto 1/ I_0$ and loss factors $1/Q_{\mathrm{J}0} \propto 1/R_\mathrm{J}$ fall perfectly into the overall trend, cf.~Fig.~\ref{fig:figure7}.
The values for the resistances $R_\mathrm{J}$ of the cavities R1 and R3 are obtained from the fits shown in Figs.~\ref{fig:figure4}\sublabel{c} and \ref{fig:figure9}\sublabel{b}, the corresponding value for R2, for which we did not extract or fit the single-junction field-modulation, is calculated from the sweetspot value of $1/Q_{\mathrm{J}0}$ shown in Fig.~\ref{fig:figure3}\sublabel{d}.
This indicates that indeed $R_\mathrm{J} \approx R_\mathrm{n}$, i.e., that the low-amplitude parallel resistance at microwave frequencies is identical to the dc resistance in the voltage state $R_\mathrm{n}$, and that maximizing $R_\mathrm{n}$ for constant $I_0$ will be a path towards reduced microwave losses and enhanced quality factors.
\begin{figure*}
	\includegraphics{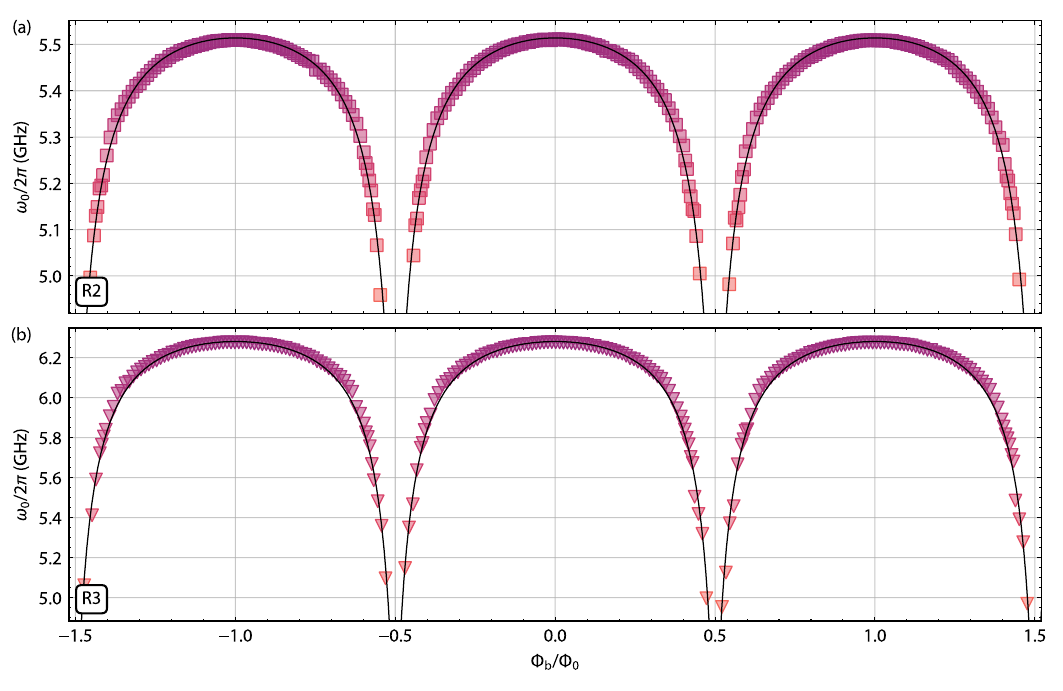}
	\vspace{-6mm}
	\titlecaption{Flux-tuning arcs of the SQUID cavities R2 and R3.}{%
		Resonance frequency $\omega_0$ vs normalized bias flux $\Phi_\mathrm{b}/\Phi_0$ for cavities~\sublabel{a}~R2 and~\sublabel{b}~R3. Symbols are data extracted from individual $S_{21}$ traces. From fits to the periodic flux modulation curves, overlaid to the data as black lines, we obtain the (mean) critical current of a single junction $I_0 = \qty{1.90}{\micro\ampere}$ in the case of R2 and $I_0 = \qty{1.51}{\micro\ampere}$ in the case of R3. Deviations between data and fit for R3 are likely due to the limitations of the used circuit model, cf.~text.
	}
	\label{fig:figure8}
\end{figure*}
%

%
\subsubsection*{K.2:~Flux-tuning arcs of R2 and R3}
%
In Fig.~\ref{fig:figure8} we show the resonance-frequency flux-tuning curves of the resonators R2 and R3, similar to the data of R1 in Fig.~\ref{fig:figure3}\sublabel{b}.
The few lowest frequency points on each arc have been extracted as the minima of the background-corrected and slightly smoothed $|S_{21}|$ instead of via the fitting routine of the resonances described in Appendix~\ref{app:fitting}, since the cavities are highly undercoupled in this regime with large $\kappa_\mathrm{int}$ and the automated fitting routine failed as a consequence.
After bJJ writing, the (sweetspot) resonance frequency of R2 has been shifted down from $\omega_{\mathrm{c}2} = 2\pi\times\qty{5.81}{\giga\hertz}$ to $2\pi\times \qty{5.51}{\giga\hertz}$ and by applying flux $\Phi_\mathrm{b}$ the resonance can be further tuned to lower values by more than $\qty{500}{\mega\hertz}$.
At even lower frequencies, the resonances are so undercoupled that a reliable determination of $\omega_0$ is not feasible.
For R3, the initial shift is from from $\omega_{\mathrm{c}3} = 2\pi\times\qty{6.727}{\giga\hertz}$ to $2\pi\times \qty{6.280}{\giga\hertz}$ and the $\Phi_\mathrm{b}$ tuning range is ${\gtrsim}\qty{1}{\giga\hertz}$.
Both SQUIDs have small screening parameters with $\beta_L = 0.066$ and $0.052$, and the maximum flux responsivities in the range of the data points are $\partial\omega_0/\partial\Phi_\mathrm{b} \approx 2\pi\times\qty{10}{\giga\hertz}/\Phi_0$ for R2 and $\partial\omega_0/\partial\Phi_\mathrm{b} \approx 2\pi\times\qty{24}{\giga\hertz}/\Phi_0$ for R3.
More details on the cavity and bJJ parameters can be found in tables~\ref{tab:table1} and~\ref{tab:table2}.
The slight mismatch between fit line and data of R3 could stem from our model reaching its limits for this cavity, more specifically to the first order Taylor approximation in Eq.~(\ref{eqn:Taylor_cot}), which requires $|\beta l - \pi/2| \ll \pi/2$, a condition not fulfilled anymore when $(\omega_\mathrm{c} - \omega_0)/\omega_\mathrm{c} \gtrsim 0.2$, which for R3 is reached at ${\sim}\qty{5.38}{\giga\hertz}$.

\begin{figure*}
	\includegraphics{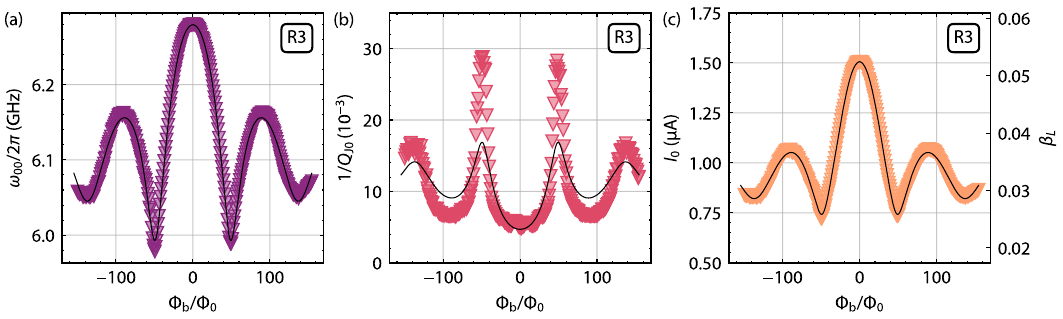}
	\vspace{-4mm}
	\titlecaption{Fraunhofer-tuning of Josephson cavity R3.}{%
		From the data shown in Fig.~\ref{fig:figure4}\sublabel{a} we can not only extract the sweetspot parameters for R1 as shown in Fig.~\ref{fig:figure4}\sublabel{b}\textendash\sublabel{d}, but also the analogous values for R3.
		\sublabel{a}~Sweetspot resonance frequency $\omega_{00}$ of R3, showing a decaying sweetspot modulation between $\qty{6.28}{\mega\hertz}$ and $\qty{5.95}{\mega\hertz}$, with clear side-maxima and minima reminiscent of a Fraunhofer-like interference pattern.
		\sublabel{b}~Junction-induced sweetspot loss factor $1/Q_\mathrm{J0}$ vs $\Phi_\mathrm{b}/\Phi_0$, showing similar but inverted features as $\omega_{00}$.
		\sublabel{c}~Critical junction current $I_0$ and screening parameter $\beta_L$ as a function of $\Phi_\mathrm{b}/\Phi_0$.
		Symbols in all panels are experimental data, the black line in \sublabel{c} is a fit to the $I_0$ data as described in Appendix~\ref{app:Fraunfit}, the black line in \sublabel{a} is the theory curve based on the fit in \sublabel{c}, and theblack line in \sublabel{b} is a fit based on the fit curve in \sublabel{c} as well as the junction resistance $R_\mathrm{J}$ as single fit parameter.
	}
	\label{fig:figure9}
\end{figure*}

\vspace{-3mm}
\subsubsection*{K.3:~Field-tuning of R3}
%
Figure~\ref{fig:figure4}\sublabel{a} contains the field-tuning data of all three cavities, but in panels \sublabel{b}\textendash\sublabel{d} we only discussed the extracted values of $\omega_{00}$, $1/Q_{\mathrm{J}0}$, and $I_0$ for R1.
In Fig.~\ref{fig:figure9} we present the anaologous data for R3.
For R2 we refrained from doing the same analysis for two reasons:~First, the resonance crosses a significant background resonance, cf.~the horizontal line at ${\sim}\qty{5.43}{\giga\hertz}$ in Fig.~\ref{fig:figure4}\sublabel{a}, which renders the automatic resonance and sweetspot detection we use for R1 and R3 non-operational, and a manual extraction would have been very time-demanding.
Secondly, we observe skewed flux tuning arcs in a considerable range around the first Fraunhofer-minima of R2, cf.~Fig.~\ref{fig:figure11}, which indicates field-dependent Josephson-diodes~\cite{Wilde2025}.
In Josephson-diodes, the sweetspot Josephson inductances are potentially decoupled from the junction critical currents due to non-sinusoidal current-phase-relations~(CPRs) and hence extracting $I_0$ from $L_{\mathrm{J}0}$ could be rather misleading.

\begin{figure}
	\includegraphics{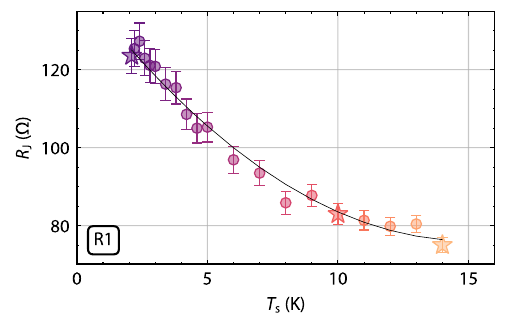}
	\vspace{-4mm}
	\titlecaption{Temperature-dependence of $R_\mathrm{J}$ in cavity R1.}{%
		Symbols represent the Josephson resistance $R_\mathrm{J}$ calculated from the experimental sweetspot quality factors $Q_{\mathrm{J}0}$ shown in Fig.~\ref{fig:figure5}\sublabel{d} using Eq.~(\ref{eqn:losses_main}); star symbols correspond to the flux arc data shown in Fig.~\ref{fig:figure5}\sublabel{a}, the line is a parabolic fit using Eq.~(\ref{eqn:RT_app}). Error bars result from the error in $1/Q_{\mathrm{J}0}$, cf.~Appendix\ref{app:errors}. The resistance decreases with increasing temperature $T_\mathrm{s}$ from ${\sim}\qty{125}{\ohm}$ at $\qty{2.1}{\kelvin}$ to ${\sim}\qty{75}{\ohm}$ at $\qty{14}{\kelvin}$.
	}
	\label{fig:figure10}
\end{figure}

Regarding the two analyzed cavities, the field-modulation of the sweetspot characteristics is similar for R1 and R3, but there are also some clear differences.
The modulation depth of $\omega_{00}$ due to the single-junction interference of R3 (${\sim}\qty{300}{\mega\hertz}$) is much smaller than the SQUID tuning of the central SQUID arcs as shown in Fig.~\ref{fig:figure8}\sublabel{b}, while for R1 the two ranges were comparable.
Correspondingly, the critical current is suppressed to only half its maximum value in the first minima, while for the bJJs of R1 the minima were at around $0.15$ of the maximum value.
Furthermore, the first side-domes reach higher values compared to the absolute maximum than in R1 and the second minima seem somewhat more strongly washed out.
These observations could indicate that the junctions do not respond to the field in sync, but that the two involved single-bJJ interference patterns differ in shape and location of the minima and side-maxima.
We cannot find a clear signature for an asymmetric SQUID from the tuning arcs around the minima or side-maxima though, such as a reduced frequency-modulation amplitude, but in particular for very small critical currents and unresolved resonance frequencies near the flux-tuning minimum quite large asymmetries might go unnoticed~\cite{Paradkar2025}.
Other potential origins for this unusual observation could be non-sinusoidal CPRs induced by the magnetic field, rendering the simple correspondence between $L_{\mathrm{J}0}$ and $I_0$ invalid, or a yet unclear impact of thermal noise currents, which are on the same order of magnitude as the critical currents in the first Fraunhofer minima.
Furthermore, for device R3 the sweetspot loss factor $1/Q_{\mathrm{J}0}$ is not described too well by the simple model Eq.~(\ref{eqn:losses_main}), cf.~Fig.~\ref{fig:figure9}\sublabel{b}.
While the fit function with constant $R_\mathrm{J}$ matches the data for $\Phi_\mathrm{b} = 0$ quite accurately, it underestimates the losses in the sharp maxima and overestimates the losses around the first side-minima.
We believe that the two effects, the unusually low modulation of the critical current with field and the deviation between theory and experiment regarding the losses, are likely related to each other.
Currently we are not able to explain the exact mechanism, but it is noteworthy that our simple theoretical treatment, in particular regarding $1/Q_{\mathrm{J}0}$, does not take into account all aspects that might play a role here.
Fluxoid quantization including the loop inductance, current-conservation in presence of parallel resistors, SQUID loop asymmetries, non-sinusoidal CPRs, and the SQUID being embedded in a wide CPW, in which the microwave current density is concentrated near the center conductor edges when it approaches the SQUID, are all factors that will refine the model.
If all these things are taken into account, the simple parallel combination of two Josephson inductances and two resistors used in the derivation of $1/Q_{\mathrm{J}0}$ likely loses validity, and only a more accurate model might be able to capture and explain the observations.

\vspace{-3mm}
\subsubsection*{K.4:~Temperature-dependence of \texorpdfstring{$R_\mathrm{J}$}{RJ} in R1}
%
In Fig.~\ref{fig:figure5}, we investigate the junction and cavity characteristics of R1 as a function of the sample temperature $T_\mathrm{s}$.
In panel \sublabel{d} we presented the temperature-dependence of the loss factor $1/Q_{\mathrm{J}0}$ and performed a fit with Eq.~(\ref{eqn:losses_main}) and a temperature-dependent $R_\mathrm{J}(T_\mathrm{s})$ as the only free parameter.
For completeness, we show both the analytically calculated $R_\mathrm{J}$ for each $1/Q_{\mathrm{J}0}$ and the fit curve in Fig.~\ref{fig:figure10}.
The analytically calculated data were obtained by inverting Eq.~(\ref{eqn:losses_main}), and the phenomenological fit function is
\begin{equation}
	R_\mathrm{J}(T_\mathrm{s}) = R_0 + r_0\left(T_\mathrm{s} - T_\mathrm{ch}\right)^2
	\label{eqn:RT_app}
\end{equation}
with $R_0$, $r_0$ and $T_\mathrm{ch}$ as the fit parameters.
In the investigated range the resistance decreases monotonously from ${\sim}\qty{125}{\ohm}$ at $\qty{2.1}{\kelvin}$ to ${\sim}\qty{75}{\ohm}$ at $\qty{14}{\kelvin}$, which is consistent with earlier results in dc experiments~\cite{Cybart2015, Couedo2020}, in which a similar increase of the resistance in low-critical-current YBCO He-bJJs with decreasing~$T_\mathrm{s}$ was observed, even with similar ratios between low-$T_\mathrm{s}$ and high-$T_\mathrm{s}$ resistances.

\vspace{-1mm}
\subsubsection*{K.5:~Josephson-diode flux-tuning in R2}
%
\begin{figure*}
	\includegraphics{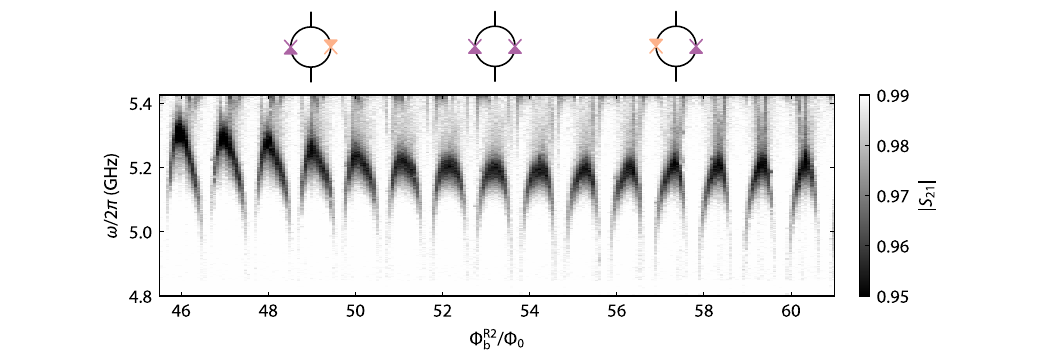}
	\vspace{-6mm}
	\titlecaption{Josephson-diode flux-tuning arcs in cavity R2.}{%
		Transmission $|S_{21}|$ vs bias flux $\Phi_\mathrm{b}^\mathrm{R2}/\Phi_0$ through the SQUID of R2 around the first Fraunhofer-diffraction minimum, cf.~Fig.~\ref{fig:figure4}\sublabel{a}. Dark regions are the absorption resonances. For the lowest flux values shown, the tuning arcs are skewed to the left, at $\Phi_\mathrm{b}/\Phi_0 \sim 53$ the tuning arcs are symmetric and unskewed again, for larger fluxes, i.e., in the right third of the figure, the tuning arcs are skewed to the right. Such skewed flux responses have recently been explained by a SQUID made of two equal-direction Josephson-diodes relative to the circulating current in the SQUID loop~\cite{Wilde2025}. Here, the three regimes would therefore correspond to the two bJJs being clockwise diodes, being opposite direction diodes (or no diodes), and being counter-clockwise diodes, respectively. Clockwise and counter-clockwise could also be vice versa, since we do not know the polarity of the magnetic field. Sketches above the three regimes schematically show one set of possible diode configurations in the cavity SQUID. A second and independent indicator for Josephson diodes is the loss of the strict periodicity of the flux modulations visible here, since Josephson diodes often have a finite phase difference as their ground state, which shifts with the external field.
	}
	\vspace{-3.5mm}
	\label{fig:figure11}
\end{figure*}
\begin{figure}
	\includegraphics{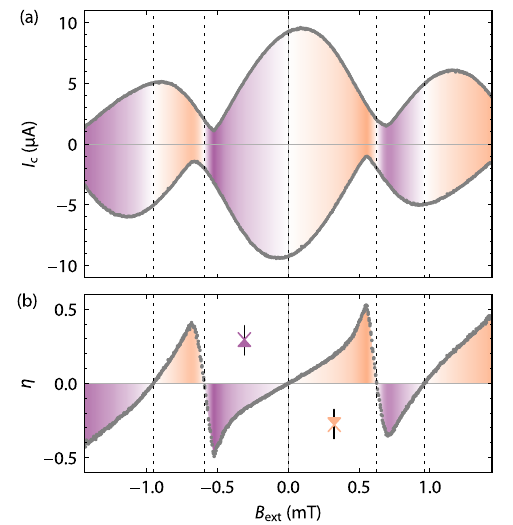}
	\vspace{-4.7mm}
	\titlecaption{A bJJ Josephson-diode with sign reversal near the first Fraunhofer minima.}{%
		 \sublabel{a}~Positive and negative critical currents $I_{\mathrm{c}+}$ and $I_{\mathrm{c}-}$ of the dc bJJ discussed in Fig.~\ref{fig:figure2}(a,~b) vs external magnetic field $B_\mathrm{ext}$. For most $B_\mathrm{ext}$ we find $I_{\mathrm{c}+} \neq - I_{\mathrm{c}-}$.
		 \sublabel{b}~Diode efficiency $\eta$ vs external magnetic field $B_\mathrm{ext}$.
		 Dashed vertical lines in both panels indicate field values for which $I_{\mathrm{c}+} = -I_{\mathrm{c}-}$, i.e., $\eta = 0$. Color shades under the curves encode the strength of the diode effect, purple for $\eta < 0$, orange for $\eta > 0$. The maximum diode effects appear around the first minima of the Fraunhofer-like diffraction pattern, with a steep transition from a maximum positive to a maximum negative diode at $B_\mathrm{ext} \sim \pm \qty{0.6}{\milli\tesla}$. Magnetic fields are manually shifted by $\qty{-13}{\micro\tesla}$ to have $I_{\mathrm{c}+} = - I_{\mathrm{c}-}$ at $B_\mathrm{ext} = 0$.
	}
	\vspace{-4mm}
	\label{fig:figure12}
\end{figure}

We observed a slight skewing of the flux tuning arcs in the zoom-window of R2 in Fig.~\ref{fig:figure4}\sublabel{a}.
This is not an optical illusion or caused by some low-resolution artifacts.
Figure~\ref{fig:figure11} shows a wider zoom of that region, i.e., around the first minimum of the Fraunhofer-diffraction pattern of R2.
Clearly, there is a skewing of the flux tuning arcs to the left for $\Phi_\mathrm{b}/\Phi_0 \sim 46$, symmetric unskewed arcs around $\Phi_\mathrm{b}/\Phi_0 \sim 53$, and skewing of the flux tuning arcs to the right for $\Phi_\mathrm{b}/\Phi_0 \sim 60$.
The skewing becomes much stronger than the one visible in Fig.~\ref{fig:figure4} when going further away from the minimum.
As has recently been reported~\cite{Wilde2025}, such skewed flux arcs are a signature of the JJs in the SQUID being Josephson diodes, i.e., being Josephson junctions with different critical currents in positive and negative current direction.
Different critical currents in the two current directions are also connected to non-sinusoidal CPRs and to different slopes of the CPR for sub-critical currents of opposite polarity.
Since the slope of the CPR is ${\propto}1/L_\mathrm{J}$, this translates to different Josephson inductances for circulating loop currents of equal magnitude but opposite polarity and as a consequence to asymmetric flux tuning characteristics.
To observe the skewing, the two bJJs in the SQUID cannot be diodes with opposite sign and equal magnitude of the diode effect (direction relative to the circulating current), but they must either have equal signs or different strengths of the diodic characteristics~(e.g.~one bJJ is a diode and the other is not).
We do not know nor can we quantitatively analyze which is the case here, but the three small SQUID schematics in Fig.~\ref{fig:figure11} show one possible configuration for all three regimes.
When the arcs tilt to the left, the diodes both point in clockwise direction, when the arcs are symmetric one diode points clockwise and the other counter-clockwise~(or both bJJs are no diodes), and when the arcs tilt to the right both diodes point counter-clockwise.
The observation that the arc skewing and the diode effect seem to be strong close to the Fraunhofer minimum, and reverse polarity when traversing the critical current minimum, is not a coincidence but is naturally explained by the asymmetric Fraunhofer-diffraction pattern of Josephson diodes, cf.~e.g.~Ref.~\cite{Schmid2025a}.
This can be clearly illustrated with the devices studied here by plotting not only the positive critical current $I_{\mathrm{c}+}$ of one of our dc junctions as we did in Fig.~\ref{fig:figure2}\sublabel{b}, but additionally its negative counterpart~$I_{\mathrm{c}-}$, cf.~Fig.~\ref{fig:figure12}.
What might have looked like a field offset at first glance is actually also a signature of this bJJ being a field-tunable Josephson diode.
The negative critical current $I_{\mathrm{c}-}$ is an almost perfectly point-reflected copy of $I_{\mathrm{c}+}$ and the minima of $-I_{\mathrm{c}-}$ and $I_{\mathrm{c}+}$ are not appearing at the same values of $B_\mathrm{ext}$.
A useful parameter to quantify the strength and the sign of the diode effect is the diode efficiency
\begin{equation}
	\eta = \frac{I_{\mathrm{c}+} - |I_{\mathrm{c}-}|}{I_{\mathrm{c}+} + |I_{\mathrm{c}-}|}
\end{equation}
which is shown in Fig.~\ref{fig:figure12}\sublabel{b}.
Clearly, the diode effect is strongest near the minima of the diffraction pattern and reverses its sign there, the transition from the positive maximum to the negative maximum spans roughly $\Delta B_\mathrm{ext} \sim \qty{0.15}{\milli\tesla}$.
These $\qty{0.15}{\milli\tesla}$ correspond to ${\sim}15\Phi_0$ in the SQUID of the microwave circuits, which supports our interpretation of the diode arcs in cavity R2, cf. Fig.~\ref{fig:figure11}.
Note, however, that this agreement is rather coincidental, since each JJ has its own strength of the diode effect and its own scaling with the field; it nevertheless demonstrates that our reasoning is sound.
The inequality of the diode effect between different bJJs is even a necessary condition for observing the diode effect in the flux tuning data, since two completely identical bJJs would always lead to symmetric arcs, independent of their diode strength, cf.~again the center schematic in Fig.~\ref{fig:figure11}.

Since skewed arcs seem potentially beneficial for maximizing the flux responsivity and to obtain bimodal relations between $\omega_0$ and other circuit parameters like responsivity or Kerr anharmonicity~\cite{Wilde2025}, it might be an interesting route to intentionally include tailored YBCO-based Josephson diodes in future devices~\cite{Schmid2025a}.
With the appropriate junction geometry, very large asymmetries can be achieved, and the yet unexplored potential of integrating Josephson diodes into microwave circuits can further be explored and evaluated, without the need for large magnetic fields as were required~in~Ref.~\cite{Wilde2025}.
Note finally that a second and independent signature of diodic-bJJ SQUIDs can also be seen in Fig.~\ref{fig:figure11}: The arcs are not strictly periodic with bias flux anymore, since a gradual shift from being non-diodic to being diodic can also gradually shift the bJJ ground state phase~\cite{Wilde2025}.
While the first arc is still positioned quite precisely at $\Phi_\mathrm{b}/\Phi_0 = 46$, the period gets slightly longer than $\Phi_0$ with increasing arc number, such that until $\Phi_\mathrm{b}/\Phi_0 = 60$ only ${\sim}13.5$ arcs are counted instead of the expected $14$.
Further to the left ($\Phi_\mathrm{b}/\Phi_0 \lesssim 40$) as well as further to the right ($\Phi_\mathrm{b}/\Phi_0 \gtrsim 65$), the periodicity is restored.
\vspace{-3mm}
\subsection{Fraunhofer-like fit function}
\label{app:Fraunfit}
\vspace{-2mm}
To fit the anomalous Fraunhofer patterns of the bJJs, cf.~Figs.~\ref{fig:figure4}\sublabel{d} and~\ref{fig:figure9}\sublabel{c}, we use the phenomenological function
\begin{align}
	I_B(\Phi_\mathrm{J}) & = I_\mathrm{off} + I_\mathrm{a}(\Phi_\mathrm{J}) \sqrt{\left(1 - g_1^2 \right)\frac{\sin\left(\left|\pi\frac{\Phi_\mathrm{J}}{\Phi_0}\right|^{g_2}\right)^2}{\left|\pi\frac{\Phi_\mathrm{J}}{\Phi_0}\right|^{2g_2}} + g_1^2}
\end{align}
with
\begin{align}
	I_\mathrm{a}(\Phi_\mathrm{J}) & = I_{\mathrm{a}0}\left(1 + g_3\left|\pi\frac{\Phi_\mathrm{J}}{\Phi_0}\right|^{2g_4}\right)^{g_5}
\end{align}
and $\Phi_\mathrm{J} = g_6\Phi_\mathrm{b}$.
All the $g_m$ with $m\in\{1, 2, 3, 4, 5, 6\}$ as well as $I_{\mathrm{a}0}$ and $I_\mathrm{off}$ are fit parameters here.
Note that this function is not based on a specific model, but merely found by testing modifications to the usual Fraunhofer interference pattern until we are able to approximately capture the experimental results.
Hence, the values of the fit parameters do not have a physical meaning.
\vspace{-3mm}
\subsection{Error bars}
\label{app:errors}
%
For all data in the manuscript without visible error bars, the standard error from the fit routine was either smaller than the data point symbol~[Figs.~\ref{fig:figure3}(b, c),~\ref{fig:figure4}(b, c d),~\ref{fig:figure5}(a, b, c),~\ref{fig:figure8},~\ref{fig:figure9}] or no error was determined~[Figs.~\ref{fig:figure1}(d, e),~\ref{fig:figure2}(a, b, c),~\ref{fig:figure3}\sublabel{a},~\ref{fig:figure4}\sublabel{a},~\ref{fig:figure11}, and~\ref{fig:figure12}].
The visible error bars for $1/Q_{\mathrm{J}0}$ and $Q_{\mathrm{J}0}$ in Figs.~\ref{fig:figure3}\sublabel{d} and \ref{fig:figure5}\sublabel{d} were determined as follows.
We calculate the loss factor via
\begin{equation}
	\frac{1}{Q_{\mathrm{J}0}} = \frac{1}{Q_{\mathrm{int}0}} - \frac{1}{Q_\mathrm{int,c}},
\end{equation}
where we get $Q_{\mathrm{int}0}$ and $Q_\mathrm{int,c}$ directly from the resonance fitting routine with and without bJJs, respectively.
Hence, the error is
\begin{equation}
	\Delta\left(\frac{1}{Q_{\mathrm{J}0}}\right) = \frac{1}{Q_{\mathrm{int}0}^2}\Delta Q_{\mathrm{int}0} + \frac{1}{Q_\mathrm{int,c}^2}\Delta Q_\mathrm{int,c}
\end{equation}
with the individual errors $\Delta Q_{\mathrm{int}0}$ and $\Delta Q_\mathrm{int,c}$.
The former, $\Delta Q_{\mathrm{int}0}$, we get directly as standard error from the fitting routine.
For the second, we assume $\Delta Q_\mathrm{int,c} = Q_\mathrm{int,c}/2$, which is much larger than the standard error from the fit.
The reason for assuming such a seemingly large error is that we also observe considerable changes of $\omega_\mathrm{c}$ and $\kappa_\mathrm{ext}$, and hence it would be quite unusual if $ Q_\mathrm{int,c}$ would remain unmodified after He irradiation and several months of time.
To be on the safe side, we estimate the possible change rather large.
Once we know the error in $1/Q_{\mathrm{J}0}$, we can also easily calculate the error of its reciprocal
\begin{equation}
	\Delta Q_{\mathrm{J}0} = \Delta\left(\frac{1}{Q_{\mathrm{J}0}}\right)Q_{\mathrm{J}0}^2 .
\end{equation}
Finally, we calculate an error for $R_\mathrm{J}$, as visible in Fig.~\ref{fig:figure10}, and for $I_\mathrm{c}R_\mathrm{n}$ as shown in Fig.~\ref{fig:figure7}.
The error in $R_\mathrm{n}$ we get as standard deviation of the mean of the multiple individual values for $R_\mathrm{n}$ extracted from the IVCs.
The error in $R_\mathrm{J}$ on the other hand, we get as
\begin{equation}
	\Delta R_\mathrm{J} = \frac{\omega_{00} L_{\mathrm{J}0}^2}{2L_\mathrm{r} + L_{\mathrm{J}0}}\Delta Q_{\mathrm{J}0}.
\end{equation}
For the error in the characteristic voltage, we use
\begin{align}
	\Delta V_\mathrm{c} & = I_\mathrm{c}\Delta R_\mathrm{n} 
\end{align}
In all figures with error bars, we only show those error bars that are larger than the symbol size.

\end{document}